\documentclass[journal]{vgtc}         

\ifpdf
  \pdfoutput=1\relax                   
  \usepackage{graphicx}                
  \DeclareGraphicsExtensions{.pdf,.png,.jpg,.jpeg} 
\else
  \usepackage{graphicx}                
  \DeclareGraphicsExtensions{.eps}     
\fi%

\graphicspath{{figures/}{pictures/}{images/}{./}} 

\usepackage{microtype}                 
\PassOptionsToPackage{warn}{textcomp}  
\usepackage{textcomp}                  
\usepackage{mathptmx}                  
\usepackage{times}                     
\usepackage{cite}                      
\usepackage{tabu}                      
\usepackage{booktabs}                  

\usepackage[hashEnumerators,smartEllipses]{markdown}

\usepackage{amsmath} 

\usepackage{xspace}
\newcommand{\etal}{et al.\@\xspace} 
\newcommand{\eg}{e.g.\@\xspace} 
\newcommand{\ie}{i.e.\@\xspace} 
\usepackage{color}
\usepackage{xcolor}
\usepackage{savesym} \savesymbol{spacing} 
\usepackage{setspace}
\usepackage{tikz}
\usepackage{colortbl}
\definecolor{Orange}{rgb}{1,0.5,0}
\definecolor{DarkGreen}{rgb}{0,0.5,0}
\definecolor{Purple}{rgb}{0.7,0,0.7}
\definecolor{Blue}{rgb}{0.2,0.2,0.8}
\definecolor{Red}{rgb}{1.0,0.0,0.0}
\definecolor{Brown}{rgb}{0.7,0.4,0.1}
\definecolor{Blue}{rgb}{0, 0, 1.}
\definecolor{Green}{rgb}{0., .6, 0.}
\definecolor{Custom}{rgb}{0.3, .1, 0.2}
\definecolor{Yellow}{rgb}{.9, .7, 0.}
\definecolor{Purple}{rgb}{.9, .1, 0.8}

\global\long\def\Matrix#1{\mathbf{#1}}
\global\long\def\Vector#1{\mathbf{#1}}

\global\long\def\Scalar#1{#1}

\global\long\def\Rotation{   \Matrix{R}}
\global\long\def\Translation{\Vector{t}}

\global\long\def\Point{\Vector{p}}
\global\long\def\Transformation{\Matrix{T}}

\global\long\def\CoordinateSymbol#1{#1}

\global\long\def\OperatorA#1#2{
{}^{\scriptscriptstyle #2}#1
}
\global\long\def\OperatorAB#1#2#3{
{}^{\scriptscriptstyle #3}{#1}_{\scriptscriptstyle #2}
}

\global\long\def\RotateAB#1#2{   \OperatorAB{\Rotation}{#1}{#2}}
\global\long\def\TranslateAB#1#2{\OperatorAB{\Translation}{#1}{#2}}
\global\long\def\TransformAB#1#2{\OperatorAB{\Transformation}{#1}{#2}}

\global\long\def\InCoordinates#1#2{\OperatorA{#2}{#1}}

\onlineid{1185}

\vgtccategory{Research}
\vgtcpapertype{algorithm/technique}

\title{Beaming Displays}

\author{Yuta Itoh, Takumi Kaminokado, and Kaan Ak\c{s}it}
\authorfooter{
\item
 Yuta Itoh is with Tokyo Institute of Technology. E-mail: yuta.itoh@c.titech.ac.jp.
\item
 Takumi Kaminokado is with Tokyo Institute of Technology. E-mail: kaminokado@ar.c.titech.ac.jp
\item
 Kaan Ak\c{s}it is with University College London. E-mail: kaksit@cs.ucl.ac.uk.
}

\shortauthortitle{Biv \MakeLowercase{\textit{et al.}}: Global Illumination for Fun and Profit}

\abstract{Existing near-eye display designs struggle to balance between multiple trade-offs such as form factor, weight, computational requirements, and battery life. These design trade-offs are major obstacles on the path towards an all-day usable near-eye display. In this work, we address these trade-offs by, paradoxically, \textit{removing the display} from near-eye displays. We present the beaming displays, a new type of near-eye display system that uses a projector and an all passive wearable headset. We modify an off-the-shelf projector with additional lenses. We install such a projector to the environment to beam images from a distance to a passive wearable headset. The beaming projection system tracks the current position of a wearable headset to project distortion-free images with correct perspectives. In our system, a wearable headset guides the beamed images to a user's retina, which are then perceived as an augmented scene within a user's field of view. 
In addition to providing the system design of the beaming display, we provide a physical prototype and show that the beaming display can provide resolutions as high as consumer-level near-eye displays. We also discuss the different aspects of the design space for our proposal.
} 


\CCScatlist{ 
 \CCScat{K.6.1}{Management of Computing and Information Systems}%
{Project and People Management}{Life Cycle};
 \CCScat{K.7.m}{The Computing Profession}{Miscellaneous}{Ethics}
}

\teaser{
  \centering
  \includegraphics[width=\linewidth]{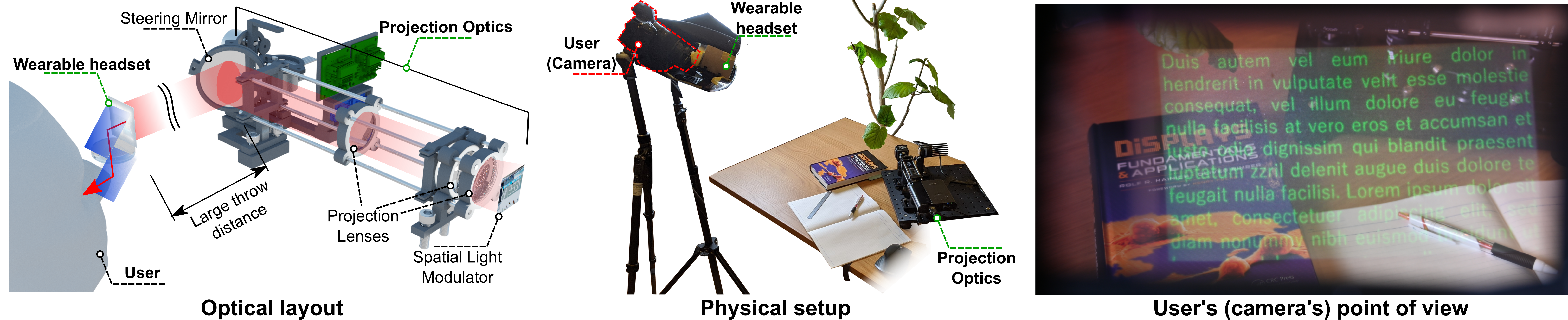}
  \caption{Beaming Displays. Left: Our new optical layout decomposes a near-eye display design into two parts, an all passive light-weight wearable headset, and a remotely located projector, this decomposition effectively avoids trade-offs between ergonomics, computational and power requirements. Middle: We build a physical setup to demonstrate the possibilities with our optical layout. Right: We show experimentally that our design supports resolutions matching a consumer level near-eye display.}
	\label{fig:teaser}
}

\vgtcinsertpkg

\begin{document}


\firstsection{Introduction}\label{sec:introduction}

\maketitle
Augmented Reality (AR) near-eye displays promise to improve our daily lives with countless applications in communication, healthcare, and manufacturing industries. 
However, there are technical challenges and obstacles for the mass adoption of AR near-eye displays.
Those technical challenges mostly lie in achieving compact form factors, while equipping an AR display with necessary optical components, sensors, power banks, and computational resources.

These functional demands in equipping AR near-eye displays with essential components usually cause a trade-off with ergonomics at the majority of designs, while leading to computational performance issues, heavy power draw, or thermal concerns. 
For example, LiKamWa et al.~\cite{likamwa2014draining} investigated thermal characteristics of consumer-grade near-eye displays and concluded that the surface temperature of such near-eye displays can exceed 39 degrees Celsius in 120 seconds, and can reach over 50 degrees Celsius after 10 minutes of usage.
Therefore, proper thermal management is critical for wearable displays as the continuous contact to hot surfaces can cause blood vessel damage even at low temperatures as 38 degrees Celsius.
Given the trade-offs and the health risks, existing solutions have to make a compromise in the design by either tethering a display to an external computer or equipping a display with a limited computational resource to turn it into a stand-alone computing unit.

We address these concerns by providing a new near-eye display design paradigm that provides an untethered and light-weight solution with promises of powerful computational resources without thermal concerns.

In this work, we redefine the design framework for see-through near-eye displays by physically separating the image generating parts from the eyepiece.
In this configuration, an image generating beaming unit beams images from a distance to a light receiving unit equipped with an eyepiece on the user's side.
Our final implementation can be described as a remotely controllable all passive wearable AR display with a light-weight body that is free from batteries or electronics that can heat up or any other active components that can pose design trade-offs related challenges in traditional display hardware.
Our main contributions can be summarized as followings:
\begin{itemize}
    \item Beaming displays. We introduce a new class of AR displays that decouple computational resources, power banks, light engines, and sensors from near-eye displays to provide an ergonomic design with light-weight and slim form-factor.
    \item Design space. We show that a beaming display can provide resolutions as high as consumer level near-eye displays. We further provide a detailed overview of the design space and the trade-offs.
    \item Prototypes. We build a set of functional prototypes, in which a custom projector and an all passive wearable headset are built. We also provide a detailed evaluation of our prototypes, and a detailed discussion on limitations of our implementation.
\end{itemize}

\section{Related work}\label{section:related_work}
A commonly accepted display taxonomy described by Bimber and Raskar~\cite{bimber2006modern} classifies AR displays in four classes, that are retinal, near-eye, hand-held, and spatial AR displays. 
We believe our method offers a new class of AR displays in this context by separating the image generating part from an actual near-eye display.
We review relevant work from each class to highlight the differences with respect to the new class introduced by our work.

\begin{figure}[t]
    \centering
    \includegraphics[width=\linewidth]{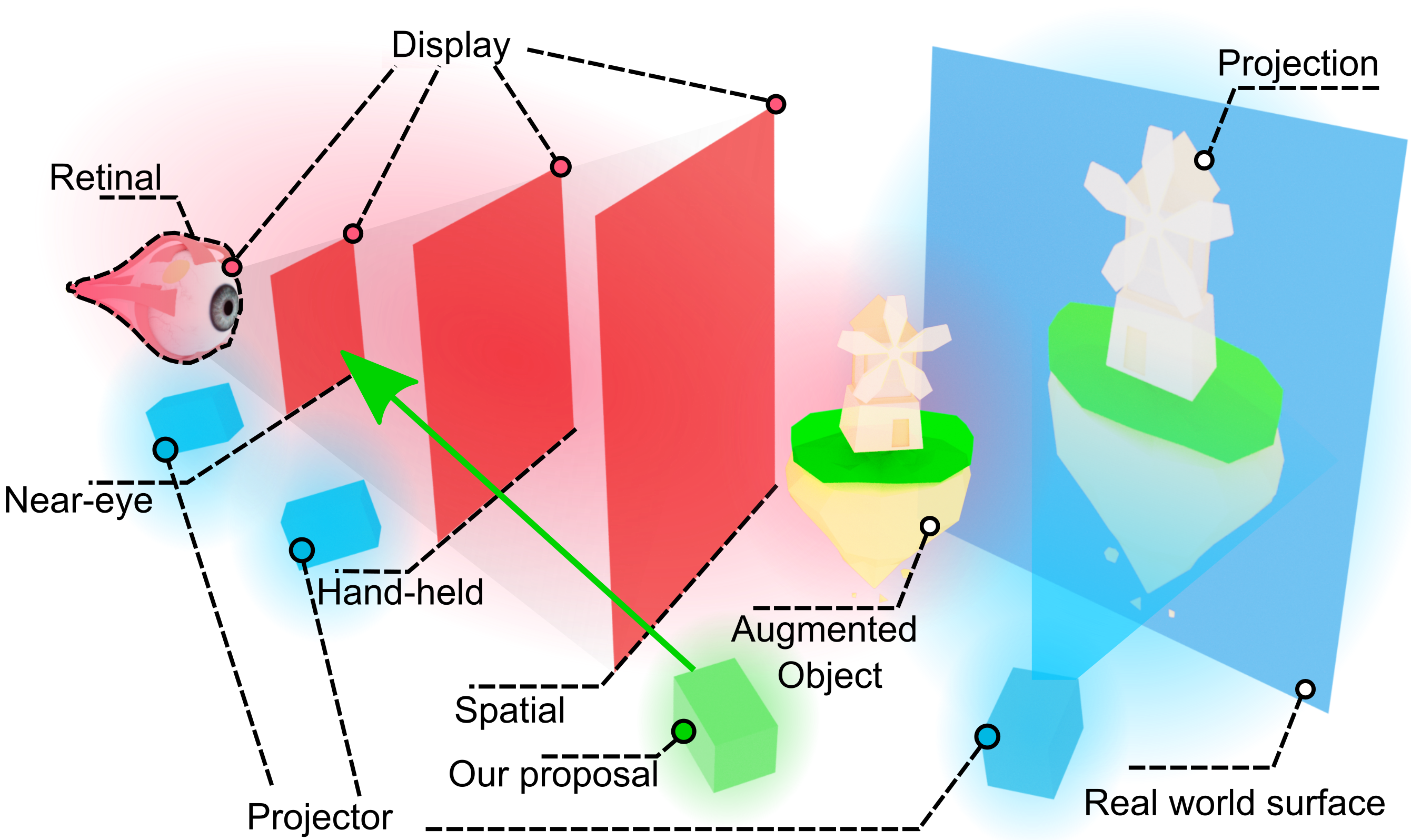}
    \caption{Types of augmented reality displays. Our proposal is distinguished as the first proposal to separate the image generating parts from an eye-piece of an augmented reality near-eye display.}
    \label{fig:type_of_ar_displays}
\end{figure}

\subsection{Retinal Augmented Reality Displays}\label{subsection:retinal_AR}
Retinal AR displays place components of an AR display in close contact with the parts of a human eye. 
In this category of AR displays, contact lens approaches are actively being explored with meta-materials~\cite{lan2019metasurfaces}, liquid crystal structures~\cite{de2013progress}, and light emitting diodes~\cite{martin2020mojo}.
A derivative of retinal AR displays is virtual retinal AR displays, where such displays typically scan a narrow laser beam on a user's retina using a Micro-Electromechanical System (MEMS) scanner~\cite{rolland2012see}.
Either retinal or virtual retinal AR displays place image generating opto-electronical parts in close contact with a user. 
To our knowledge, our proposal fundamentally differs as there has not been a passive version of such displays where the image is beamed to a passive contact lens from a large distance.

\subsection{Near-eye Augmented Reality Displays}\label{subsection:near_eye_AR}
Near-eye AR displays~\cite{sutherland1968head} promise an AR display in the form of a pair of see-through glasses. 
In the recent years, accommodation-supporting near-eye AR displays~\cite{kuo2020high}, foveated near-eye AR displays~\cite{akcsit2019manufacturing}, vision assisting near-eye AR displays~\cite{langlotz2018chroma}, occlusion capable AR displays~\cite{hamasaki2019varifocal}, and color-subtracting AR displays~\cite{kaminokado2020stainedview} have garnered interest in the research community.
For curious readers, we leave a detailed survey on near-eye AR displays by Koulieris et al.~\cite{Koulieris2019}. 
An extension to near-eye AR displays is head-mounted projection displays, in which real-world surfaces are used to project images~\cite{kade2015head}, while a user is equipped with a wearable projector. 
Our proposal is complementary to the existing types of near-eye AR display designs and promises to improved form factors while decreasing the overall weight of existing near-eye AR displays. 
Our proposal separates all the active parts in near-eye AR displays from a user, therefore not only promises to improve ergonomics but also promises to enable larger computational resources without any heating or power-related issues.

\subsection{Hand-held Augmented Reality Displays}\label{subsection:hand_held_AR}
Hand-held AR displays are a very common form of AR display systems. 
One of the earliest hand-held AR display prototypes was from the late 90s, a video see-through system using a palmtop monitor connected to an external computer~\cite{Rekimoto1995}.
In the current day, this trend is followed by smartphones, and there are AR development frameworks~\cite{herpich2017comparative} dedicated to AR application development in smartphones.
Like the early versions from the 90s, all of the current day hand-held AR applications with hand-held displays are based on video see-through approaches. 

Head-up displays are a form of augmented reality displays that are within the arm's length reach. Head-up displays contains projectors, and passive see-through projection screens that can either base on microlenses~\cite{hedili2013microlens}, retro-reflectors~\cite{soomro2016design}, cholesteric liquid crystal dots~\cite{yamamoto201616}, and nano particles~\cite{hsu2014transparent}. 
These projection systems with see-through passive screens are closely related to our approach, however hand-held AR approaches are fundamentally the opposite of our system as users of hand-held AR systems are in close contact with active components, while users are away from passive parts of the display system, and the projection screens.

\subsection{Spatial Augmented Reality Displays}\label{subsection:spatial_AR}
Spatial AR displays can be wall-like or table-like displays, which can either be in the form of see-through light-emitting transparent display surfaces or see-through projection screens.
While the head-up display screen approaches~\cite{hedili2013microlens,soomro2016design,hsu2014transparent,yamamoto201616} described in the Sec.~\ref{subsection:hand_held_AR} can in principle scale to spatial AR applications, researchers also relied on precision projection mapping techniques to augment a scene with virtual objects by using opaque projection screens~\cite{benko2014dyadic} or by using moving objects as projection screens in the physical space~\cite{okumura2012lumipen,mikawa2018variolight}.
Our approach shares similarities with the work by Okumura et al.~\cite{okumura2012lumipen} and Mikawa \etal~\cite{mikawa2018variolight}, however with one major difference, a projection screen is much smaller in size and that projection screen is part of all passive near-eye optics in our proposal.

It should also be noted that spatial AR displays require the physical display to be as large as a wall or a table, while our approach can provide a large field of view (FoV) without requiring any large space.

Bokode~\cite{mohan2009bokode} can be seen as a hybrid of a retinal and a spatial AR display. It is similar to our concept in terms of sending an image into the user's eye directly. The concept is, however, not designed for a dynamic display and the pinhole optics inevitably causes an extremely small eyebox and the FoV is bounded by the iris size.

\global\long\def\worldCoordinate{\CoordinateSymbol{W}}
\global\long\def\hmdHousingCoordinate{\CoordinateSymbol{H}}
\global\long\def\markerCoordinate{\CoordinateSymbol{M}}
\global\long\def\screenCoordinate{\CoordinateSymbol{S}}
\global\long\def\screenCoordinateAlt{{\screenCoordinate^\prime}}
\global\long\def\cameraCoordinate{\CoordinateSymbol{C}}
\global\long\def\projectorCoordinate{\CoordinateSymbol{P}}
\global\long\def\TransFromWorldToProjector{\TransformAB{\worldCoordinate}{\ProjectorCoordinate}}
\global\long\def\CoordinatesA{\CoordinateSymbol A}
\global\long\def\CoordinatesB{\CoordinateSymbol B}
\global\long\def\CoordinatesC{\CoordinateSymbol C}
\global\long\def\PointInA{\InCoordinates{\CoordinatesA}{\Point}}
\global\long\def\PointInB{\InCoordinates{\CoordinatesB}{\Point}}
\global\long\def\TransformAtoB{\TransformAB{\CoordinatesA}{\CoordinatesB}}
\global\long\def\RotateAtoB{\RotateAB{\CoordinatesA}{\CoordinatesB}}
\global\long\def\TranslateAtoB{\TranslateAB{\CoordinatesA}{\CoordinatesB}}
\global\long\def\TranslateBtoC{\TranslateAB{\CoordinatesB}{\CoordinatesC}}
\global\long\def\TranslateAtoC{\TranslateAB{\CoordinatesA}{\CoordinatesC}}

\global\long\def\TransformProjectorToScreenAlt{\TransformAB{\projectorCoordinate}{\screenCoordinateAlt}}
\global\long\def\TransformWorldToScreenAlt{\TransformAB{\worldCoordinate}{\screenCoordinateAlt}}
\global\long\def\TransformWorldToScreen{\TransformAB{\worldCoordinate}{\screenCoordinate}}
\global\long\def\TransformHMDToScreen{\TransformAB{\hmdHousingCoordinate}{\screenCoordinate}}
\global\long\def\TransformWorldToProjector{\TransformAB{\worldCoordinate}{\projectorCoordinate}}

\global\long\def\screenCoordinateAltBase{\overline{{\screenCoordinate^\prime}}}
\global\long\def\TransformProjectorToScreenAltBase{\TransformAB{\projectorCoordinate}{\screenCoordinateAltBase}}
\global\long\def\RotateProjectorToScreenAltBase{\RotateAB{\projectorCoordinate}{\screenCoordinateAltBase}}
\global\long\def\TranslateProjectorToScreenAltBase{\TranslateAB{\projectorCoordinate}{\screenCoordinateAltBase}}
\global\long\def\BSMCoordinate{\CoordinateSymbol{B}}
\global\long\def\BSMCoordinateShifted{\CoordinateSymbol{B^\prime}}
\global\long\def\TransformProjectorToBSM{\TransformAB{\projectorCoordinate}{\BSMCoordinate}}

\global\long\def\RotateProjectorToBSM{\RotateAB{\projectorCoordinate}{\BSMCoordinate}}
\global\long\def\TranslateProjectorToBSM{\TranslateAB{\projectorCoordinate}{\BSMCoordinate}}
\global\long\def\RotateProjectorToBSMShifted{\RotateAB{\projectorCoordinate}{\BSMCoordinateShifted}}
\global\long\def\TranslateProjectorToBSMShifted{\TranslateAB{\projectorCoordinate}{\BSMCoordinateShifted}}

\global\long\def\TransformProjectorToScreenAlt{\TransformAB{\projectorCoordinate}{\screenCoordinateAlt}}
\global\long\def\TransformProjectorToBSMShifted{\TransformAB{\projectorCoordinate}{\BSMCoordinateShifted}}
\global\long\def\MirrorMatrix{\Matrix{M}}

\global\long\def\RotateProjectorToScreenAlt{\RotateAB{\projectorCoordinate}{\screenCoordinateAlt}}
\global\long\def\TranslateProjectorToScreenAlt{\TranslateAB{\projectorCoordinate}{\screenCoordinateAlt}}
\global\long\def\TranslateProjectorToScreen{\TranslateAB{\projectorCoordinate}{\screenCoordinate}}
\global\long\def\normalVector{\Vector{n}}
\global\long\def\distanceProjectorAndScreenAltBase{d}
\global\long\def\distanceProjectorAndBSMShifted{d^\prime}

\global\long\def\MirrorAngle{\theta}
\global\long\def\MirrorAngleOne{\theta}
\global\long\def\MirrorAngleTwo{\phi}
\global\long\def\RotationMirror{\widetilde{\Rotation}}

\global\long\def\FocalLength{\Scalar f}
\global\long\def\Distance{\Scalar{d}}
\global\long\def\Depth{\Scalar{z}}
\global\long\def\Dioptry{\Scalar{d}}
\global\long\def\IndexForEye{\mathrm{E}}
\global\long\def\IndexForSLM{\mathrm{L}}
\global\long\def\IndexForObject{\mathrm{O}}
\global\long\def\FocalLengthEye{\OperatorA{\FocalLength}{\IndexForEye}}
\global\long\def\FocalLengthCombined{\FocalLength}
\global\long\def\LensSpacing{\Distance}
\global\long\def\LensSpacingTwo{\Distance'}
\global\long\def\PrincipalPlaneDistance{\Scalar p}
\global\long\def\PrincipalPlaneDistanceSLM{\OperatorA{\PrincipalPlaneDistance}{\IndexForSLM}}
\global\long\def\PrincipalPlaneDistanceEye{\OperatorA{\PrincipalPlaneDistance}{\IndexForEye}}
\global\long\def\DistanceToObject{\OperatorA{\Distance'}{\IndexForObject}}
\global\long\def\DistanceToObjectTwo{\OperatorA{\Distance''}{\IndexForObject}}
\global\long\def\DistanceToEye{\OperatorA{\Distance''}{\IndexForEye}}
\global\long\def\DistanceToObjectOrig{\OperatorA{\Distance}{\IndexForObject}}
\global\long\def\DistanceToEyeOrig{\OperatorA{\Distance}{\IndexForEye}}
\global\long\def\DistanceDiff{{\scriptstyle \Delta}\Scalar d}
\global\long\def\Angle{u}
\global\long\def\AngleEye{\OperatorA{\Angle}{\IndexForEye}}
\global\long\def\AngleObject{\OperatorA{\Angle}{\IndexForObject}}
\global\long\def\SpaceMatrix{\Matrix S}
\global\long\def\LensMatrix{\Matrix L}
\global\long\def\SpaceMatrixFunc#1{\SpaceMatrix({\scriptstyle #1})}
\global\long\def\LensMatrixFunc#1{\LensMatrix({\scriptstyle #1})}
\global\long\def\RayTransferMatrix{\Matrix M}

\global\long\def\RayPositionEye{\OperatorA{\RayPosition}{\IndexForEye}}
\global\long\def\RayAngleEye{\OperatorA{\RayAngle}{\IndexForEye}}
\global\long\def\RayEye{\OperatorA{\Ray}{\IndexForEye}}
\global\long\def\RayPositionObject{\OperatorA{\RayPosition}{\IndexForObject}}
\global\long\def\RayAngleObject{\OperatorA{\RayAngle}{\IndexForObject}}
\global\long\def\RayObject{\OperatorA{\Ray}{\IndexForObject}}
\global\long\def\RTMatrix{\Matrix M}
\global\long\def\RTMatrixOne{\OperatorA{\RTMatrix}1}
\global\long\def\RTMatrixTwo{\OperatorA{\RTMatrix}2}
\global\long\def\RTMatrixThree{\OperatorA{\RTMatrix}3}

\global\long\def\ImagePlaneDepthOriginal{\OperatorA{\Depth}{0}}
\global\long\def\ImagePlaneDepthTarget{\Depth}
\global\long\def\Lens{\mathrm{L}}
\global\long\def\LensDistanceOne{\OperatorA{\Distance}{\Lens}}
\global\long\def\LensMargin{\Distance^{\prime}}
\global\long\def\MagnificationRate{s}
\global\long\def\FocalLengthSLM{\OperatorA{\FocalLength}{\IndexForSLM}}
\global\long\def\FocalLengthSLMTwo{\OperatorA{\FocalLength}{\IndexForSLM'}}

\global\long\def\TransformProjectorToScreen{\TransformAB{\projectorCoordinate}{\screenCoordinate}}
\global\long\def\TransformWorldToScreen{\TransformAB{\worldCoordinate}{\screenCoordinate}}
\global\long\def\TransformWorldToScreen{\TransformAB{\worldCoordinate}{\screenCoordinate}}
\global\long\def\TransformWorldToHMD{\TransformAB{\worldCoordinate}{\hmdHousingCoordinate}}

\section{Beaming Displays}\label{sec:system-design}
A beaming display contains two main building blocks: a steering projector and a passive wearable headset.
Images generated by a steerable projector are beamed to a wearable headset, in which wearable headset is an all passive glasses dedicated to relaying beamed images to a user's eyes. The steering projector may require a tracking unit as a submodule so that it can localize the position and the orientation of a wearable headset relative to the steering projector.
In the following sections, we will discuss all three building blocks in more detail.

\subsection{Steering Projector}\label{subsec:steering_projector}
\begin{figure}[t]
    \centering
    \includegraphics[width=0.9\linewidth]{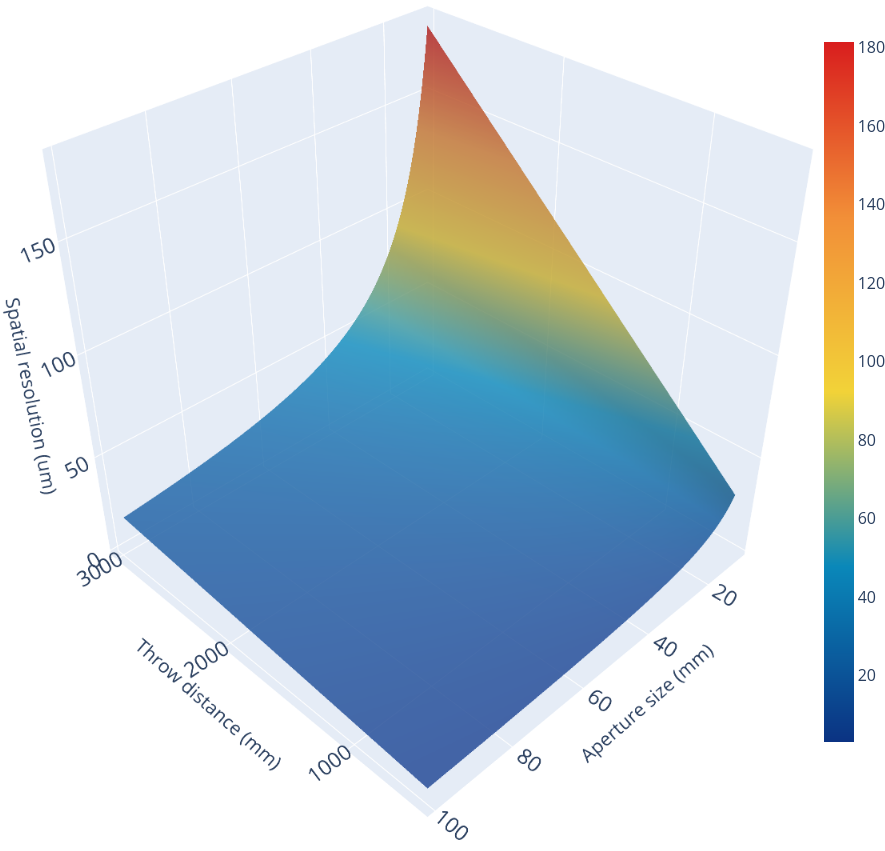}
    \vspace{-4mm}
    \caption{Rayleigh resolution analysis. Theoretical spot sizes are calculated in micro meters using Rayleigh resolution criteria over a range of aperture sizes and throw distances. Smaller the spatial resolution better the image quality.}
    \label{fig:rayleigh_resolution}
\end{figure}

The layout of a beaming display, as shown in Fig.~\ref{fig:teaser}, consists of a steering projector to project images to a wearable headset.
Off-the-shelf projectors lack the ability to steer projected images towards a certain direction and typically projects larger images than a beaming display would require.
Therefore, a dedicated projection optics with a steering mechanism is required to design a steering projector.
An optical system of a steering projector can be best described as a 4F imaging system, a lens system with two lenses. 
According to the work by Foreman et al.~\cite{foreman2011computational}, 4F imaging systems are known to suffer less from optical distortions making them an idle candidate for our projector design.
\begin{figure}[t]
    \centering
    \includegraphics[width=\linewidth]{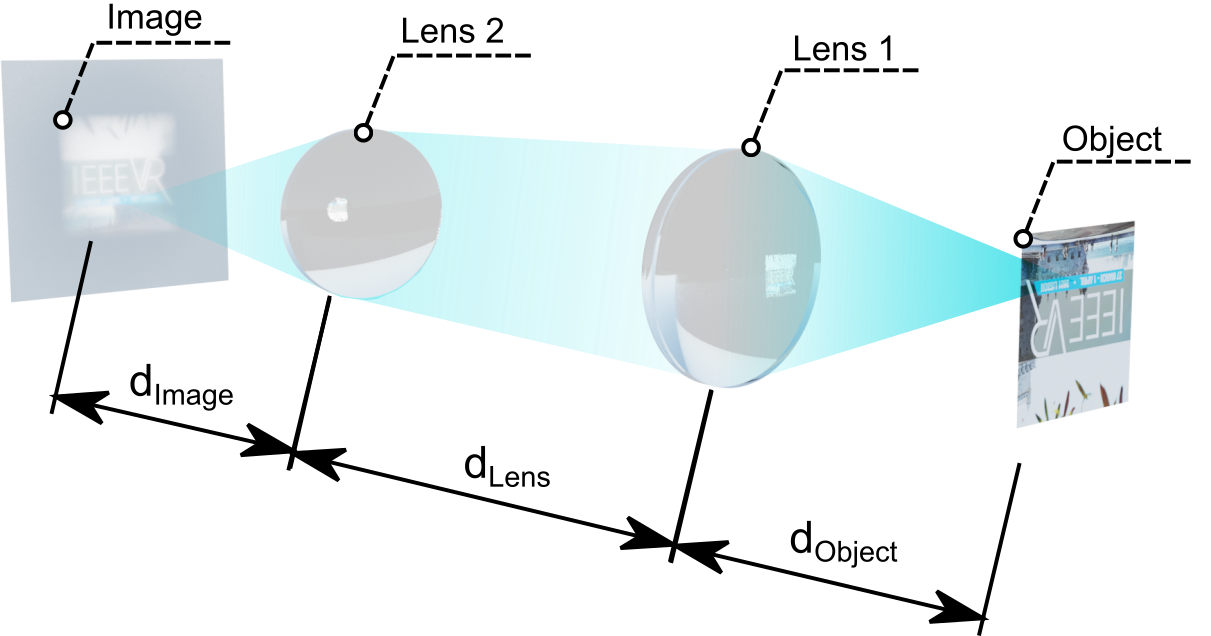}
    \vspace{-4mm}
    \caption{A 4F imaging system. An image of an object is relayed to a diffusive surface using two lenses. A beaming display's projector bases on a 4F optical system, where the throw distance of $d_{image}$ is typically in the range of $50-200$~cm. Here the object depicts a spatial light modulator, and the diffusive screen is the projection screen on a wearable headset.}
    \label{fig:system_4f}
\end{figure}
In a typical 4F system, an image of an object at object plane is relayed to an image plane as sketched in Fig.~\ref{fig:system_4f}. 
Note that Fig.~\ref{fig:system_4f} is a simplified optical layout, both the first and second lenses can be multi-element lens systems, for simplicity, our description assumes those lenses as two separate single lenses.
In a steering projector, images at the object plane are generated using an SLM, whereas an image at the image plane is an inverted copy of the image at the object plane with desired dimensions.
The first lens in Fig.~\ref{fig:system_4f} collimates the beam approaching from an SLM, while the second lens is separated by $d_{lens}$ distance in Fig.~\ref{fig:system_4f} projects an image with a throw distance of $d_{image}$, which is the effective focal length of the second lens.
Separation in between the first and the second lens $d_{lens}=f_1+f_2$ is typically chosen as the sum of focal lengths of both lenses.
The throw distance determines the resolution characteristics of projected images, and can be calculated using commonly used Rayleigh's resolution criterion,
\begin{equation}
\Delta \ell = 1.22 \frac{ d_{image} \lambda}{D},
\label{equ:rayleigh_resolution}
\end{equation}
where D is the effective aperture size of the second lens, and $\lambda$ is the wavelength of the light. 
Using Rayleigh resolution, we calculate theoretical spot size limits and compiled it as in Fig.~\ref{fig:rayleigh_resolution} with respect to a range of aperture size and throw distances.
Rayleigh resolution suggests that with throw distances as short as $50-100$~cm and with aperture size in the range of $4-6$~cm, pixel sizes as small as $10-20$~$\mu$m is possible, which is $2-3$x smaller than a pixel size of today's smartphone displays ($40-60$~um).
It should be highlighted here smaller the pixel sizes better the image quality perceived by a user.

A user can be located in different locations with respect to a steering projector, therefore the effective focal length of the second lens has to dynamically change, and with a separate mechanism, images have to be steered toward different locations.
In the projection mapping researches, focusing a projector to an object moving in depth is tackled by using focus-tunable dynamic lenses~\cite{iwai2015extended,wang2020high}, which leads to a considerable depth of fields in projection systems when the focal length is continuously swept rapidly in such lenses. 
Following the existing projection mapping research, we envision lens two in Fig.~\ref{fig:system_4f} as a focus-tunable lens. 
We also propose to install a scanning mirror right after the second lens to steer the images to different locations in space, as shown in our optical layout in Fig.~\ref{fig:teaser}.

\subsubsection{Tracking the headset}\label{subsec:tracking_unit}

\begin{figure}[t]
    \centering
    \includegraphics[width=\linewidth]{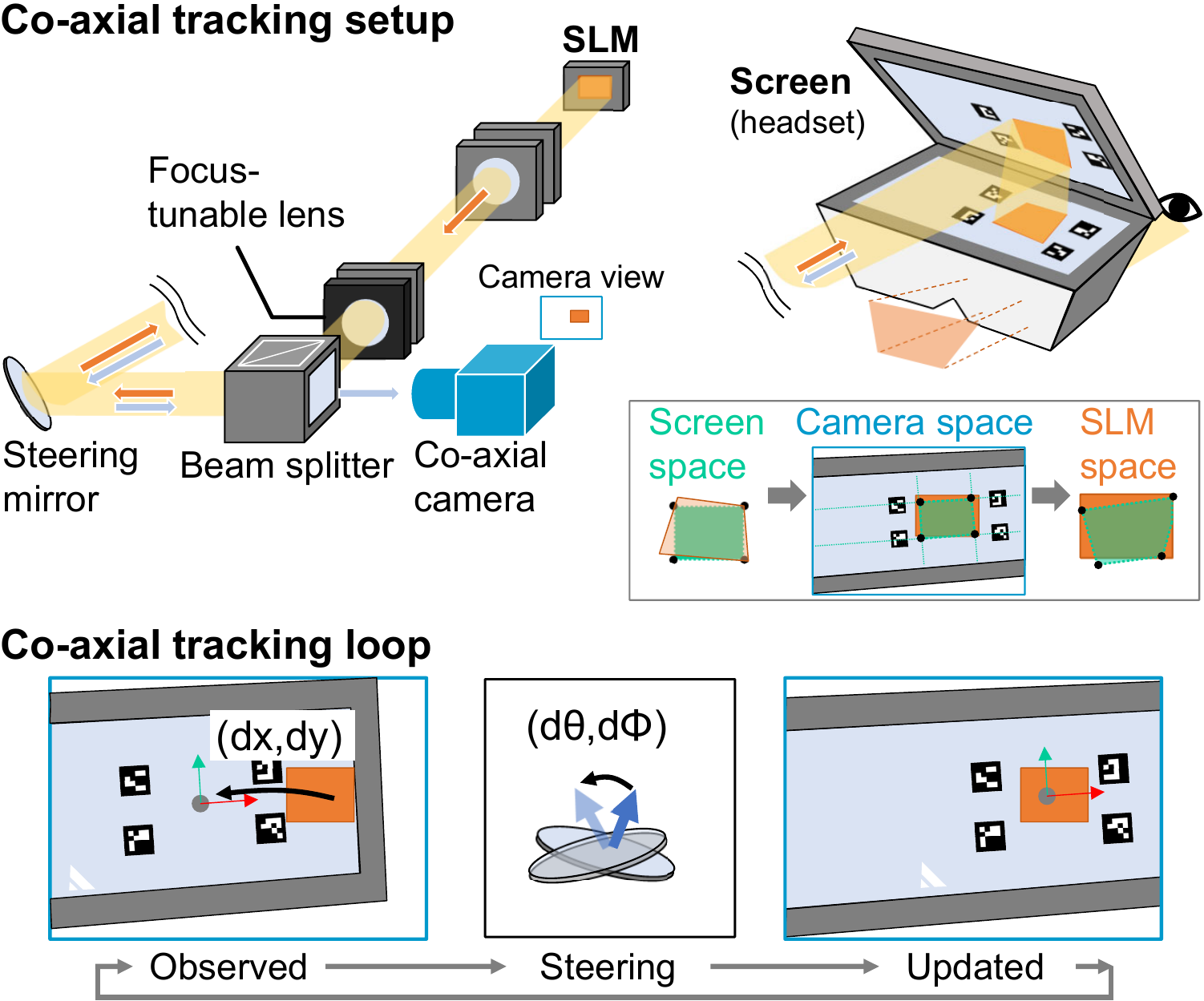}
    \caption{Top: Tracking setup. The system employs a co-axial scene camera that shares the FoV of the projection area. The tracking gives an image warping from the screen coordinates to the SLM image space. Bottom: Mirror steering. Schematic overview of the mirror steering with marker tracking by the co-axial camera in the steering projector.}
    \label{fig:system-coordinates}
\end{figure}

A steering projector requires the relative pose of the headset so that the projector can steer the mirror to cast the image at the target screen correctly.
More specifically, a tracking unit should provide the 6DoF pose of the screen coordinates toward the projector.
Based on the pose information, the steering projector then computes the projection angles $(\MirrorAngleOne,\MirrorAngleTwo)$ of a steering mirror and the focus $\FocalLength$ of a focus-tunable lens.
The design of the tracking unit could be up to use cases, and typical options are either outside-in tracking~\cite{benko2015fovear} installed in the space or inside-out tracking integrated to the projector~\cite{narita2016dynamic, mikawa2018variolight}. We elaborate our tracking implementation in Sec.~\ref{subsec:implementation:tracking_unit}.

Given the system is being tracked, we first calibrate the camera and projector offline to store the static 2D image mapping from the camera to the projector. 

When the beaming display is already steered correctly, what we need is to compute the screen shape from the camera image via marker detection.
Combined with the static projector-camera mapping, we can then compute the target image for the screen (Fig.~\ref{fig:system-coordinates} Top, bottom right).

When the projection is off from the screen area, we need to steer the mirror (Fig.~\ref{fig:system-coordinates} Bottom). By the marker detection, the scene camera obtains an offset from the center of projection to the screen center $(dx, dy)$ in the viewing direction.
Based on the offset information, the mirror can compute additional steering angles $(d\MirrorAngleOne,d\MirrorAngleTwo)$, accordingly.
Finally, a focus-tunable lens can dynamically change focus with a look-up table by using an estimation of the throw distance from detected marker positions.

\subsubsection{Tracking requirement}\label{subsec:tracking_requirement}
\begin{figure}[t]
    \centering
    \includegraphics[width=0.8\linewidth]{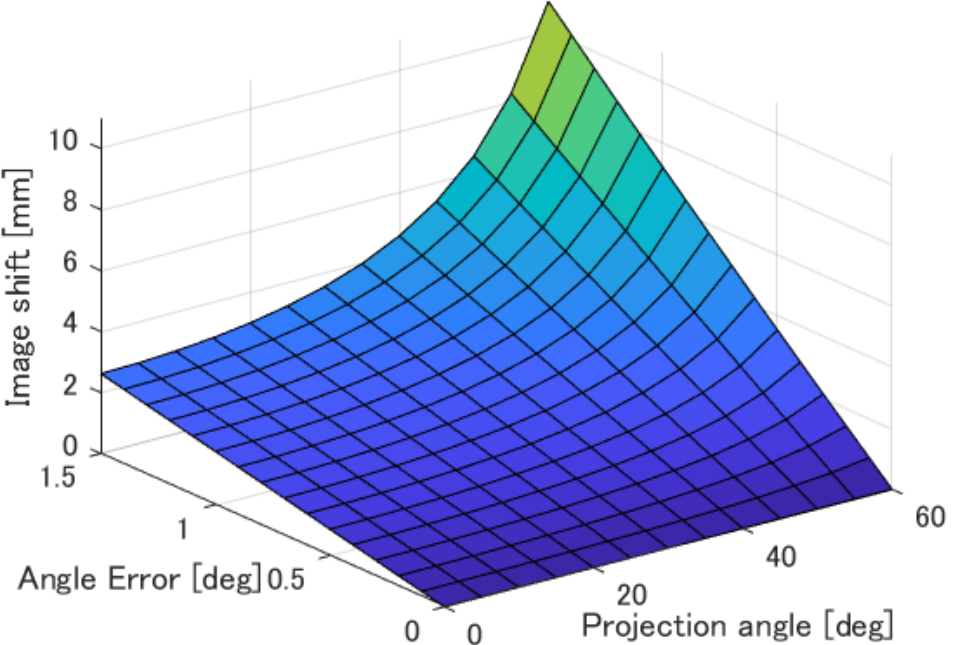}
    \caption{Steering error analysis. An image projected on the screen gets shifted by the steering angle error. We assume the projection distance at $1$m. The base projection angle amplifies the effect of the steering error.}
    \label{fig:angle-error_analysis}
\end{figure}

An important design factor for tracking is the relationship between the control accuracy of the mirror and the resulting misalignment of the projected image.
In the beam display, the screen on the headset is separated from the projection system. The projected image is also magnified by the beam combining optics. Therefore, even a small error of the mirror may cause a large misalignment in the projected image.
We analyze the error.
Given a projection distance $z$ and the steering angle $\theta$, the center of the projection may shift by 
\begin{equation}
z (\tan(\theta+\Delta\theta)-\tan(\theta)) \approx z*\Delta\theta/\cos^2\theta,
\end{equation}
with a small angular error $\Delta\theta$. Thus, the image shift corresponds to both $\theta$ and $\Delta\theta$. 
To illustrate the nature of the error, if $z=1$, $\theta=45$deg, and $\Delta\theta=0.1$deg, then it gives an error of $3.5$mm. In our prototype, which we will introduce in Sec.~\ref{sec:implementation}, the projected image height is about $20$mm, so this mm-scale shift still causes $17.5$\% of the image shift in height. As shown in Fig.~\ref{fig:angle-error_analysis}, the error increases when the original projection angle increases.

\subsection{Passive wearable headset}\label{subsec:passive_wearable_headset}
\begin{figure}[t]
    \centering
    \includegraphics[width=0.55\linewidth]{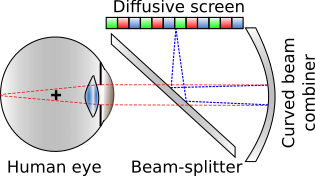}
    \vspace{-4mm}
    \caption{Birdbath optics. We choose to use birdbath optics for our passive wearable headset, which uses a beam splitter and a curved beam combiner with a screen.}
    \label{fig:birdbath}
\end{figure}

Images from the steerable projector are projected on a diffusive screen that diffuses light in the direction of the projection. 
However, simply placing that diffusive screen in front of a user's eye would lead to a blurry image, while blocking the entire FoV of that user.
Like many other AR near-eye display systems, the images on that diffusive screen has to be relayed to a user's eye in a way that generates virtual images at some distance from a user so that the user can focus and perceive sharp augmented images within the FoV.
As described in work by Bernard Kress~\cite{kress2019optical}, relaying images from a screen to a user's eyes can be done in many different ways, most traditional ones are based on using optical waveguides or bird-bath optics.
For the sake of simplicity, we base our design for an all-passive wearable headset to the most common optical layout of bird-bath optics. In this optics, a screen is relayed to a user's eyes by using a beam splitter that maintains real-world view and a curved beam combiner that helps in the generation of virtual images as sketched in Fig.~\ref{fig:birdbath}. 
Our optical layout shown in Fig.~\ref{fig:teaser} Left depicts a variant of bird-bath optics with a lens and a beam splitter.
We highlight the fact that other relaying techniques in near-eye display optics domain~\cite{kress2019optical} are, in principle, compatible with our beaming display design.

\section{Implementation}\label{sec:implementation}
We demonstrate our prototype shown in Fig.~\ref{fig:prototype}.
Our prototype contains a wearable passive headset and a steering projector with an integrated inside-out tracking unit.

\subsection{Steering projector} 
Our steering projector uses a Digital Micromirror Device (DMD) as an SLM.
We harvest this DMD from an Acer C200 projector.
This specific projector uses Light Emitting Diodes (LEDs) as the illumination, supports a native resolution of 854 by 480 pixels, and provides 200~lm brightness with a 2000 to 1 contrast ratio.
We remove the built-in projection optics of the projector to build an imaging system that can image the DMD to the desired throw distance. 
The same imaging system resizes the image of the DMD to match the dimensions of a diffusive screen on a wearable headset.

\begin{figure*}[hbt!]
    \centering
    \includegraphics[width=0.7\linewidth]{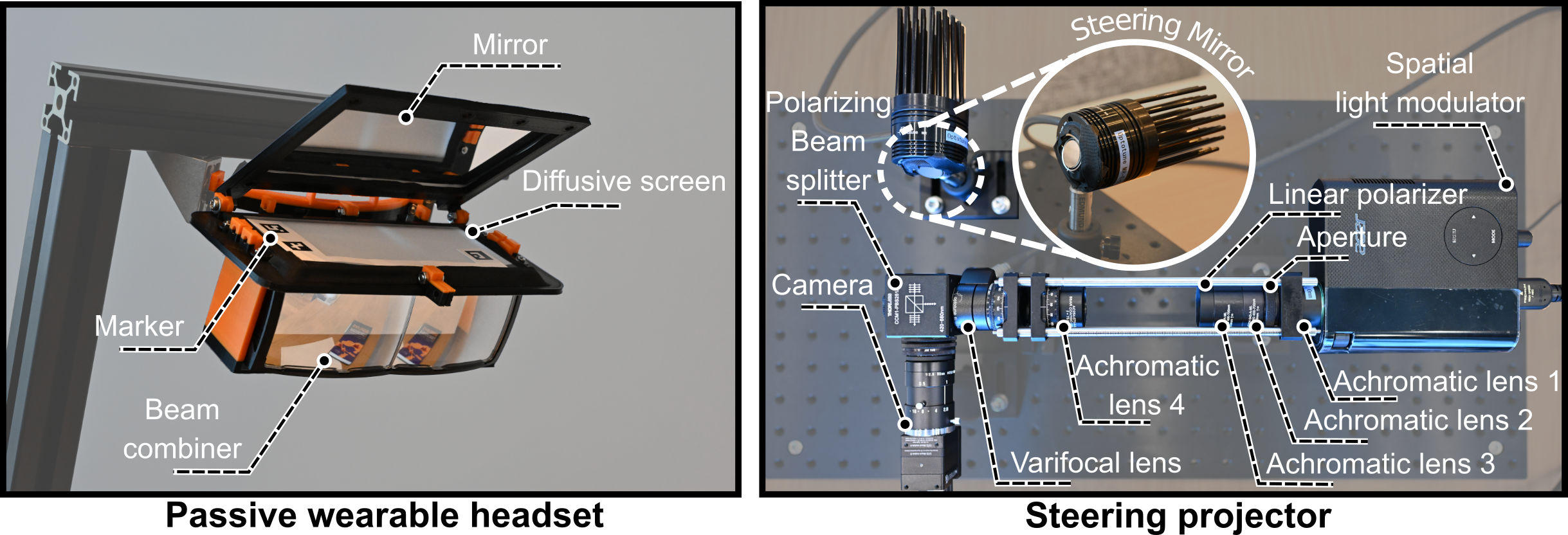}
    \vspace{-4mm}
    \caption{Beaming display prototype. Left: Photograph of our all passive wearable headset based on bird-bath optics and equipped with visible markers for tracking. Right: Photograph of our steering projector with custom varifocal optics and beam steering mirrors. Our tracking camera is co-axial to our steering projector and captures images through the steering mirror.}
    \label{fig:prototype}
\end{figure*}

As shown in the right portion of Fig.~\ref{fig:prototype}, our imaging system uses four achromatic lenses followed by a focus-tunable (varifocal) lens to support throw distances between 50 cms to 200 cms.
The achromatic lenses used in our imaging system from DMD to a varifocal lens are as follows: an AC254-045-A-ML, another AC254-045-A-ML, an AC254-030-A-ML, and an AC254-075-A-ML from Thorlabs.
In between first and second achromatic lenses, we use a ring actuated iris diaphragm aperture, Thorlabs SM1D12D, to avoid total internal reflections in our lens system caused by light rays with large angles approaching from our projector's illumination.
A linear polarizer, Thorlabs LPVISE050-A, is placed after the third achromatic lens to be able to split the optical path with a polarizing beam splitter at a later stage.

Splitting the optical path helps us to accommodate a camera for the tracking unit without interfering with the optical system.
At the end of the achromatic lenses, we place a varifocal lens of Optotune EL-10-30-Ci.
This varifocal lens can vary its focus from -1.5~diopters to 3.5~diopters, and has an effective circular aperture size of 10~mm while providing response times smaller than 2.5~ms.
Our varifocal lens is followed by a Thorlabs CCM1-PBS251/M polarizing beam splitter.

The final component in our imaging system is a steering mirror, Optotune MR-E-2 silver coated mirror, with an effective circular aperture of 15~mm, and a step resolution of 22~$\mu$rad.

The mirror supports several control protocols. For easy connection, serial communication via USB is available. However, due to the specification of the API and the overhead of the serial communication, the update rate of set a new mirror angle is limited to about 16 fps, which is not suitable for our application. For fast control, the Serial Peripheral Interface (SPI) protocol is available for direct communication with the driver board. Our implementation thus employs the SPI protocol. Specifically, we use an ARM-based development board, Nucleo-144 STM32F746ZGT6, to generate SPI signals, and the host PC controls the mirror through this board via Ethernet UDP connection written in C++. In this setup, the maximum command update rate of the mirror can easily reach more than 1,000 fps. Note that, however, the mirror step settling time is physically bounded (e.g. 2msec for 0.1deg. and 12msec for 20deg.).

\subsubsection{Tracking unit}\label{subsec:implementation:tracking_unit}
Similar to an existing dynamic projector system~\cite{mikawa2018variolight}, we employ an inside-out tracking by implementing a scene camera that is co-axial to the projection path (Fig.~\ref{fig:system-coordinates} Top).
Thanks to the co-axial design, the camera and the projector share the same FoV and the projection area stays static in the camera image. The camera images the scene that the steering mirror points at any time.
Once the camera detects the screen, we can compute the mapping from the projector to the screen.

The tracking unit of our prototype is integrated into our steering projector using the Thorlabs CCM1-PBS251/M polarizing beam splitter in the steering projector. We add a Ximea MC031CG-SY-UB camera with $50$~mm Tamron M118FM50 C-mount lens, which can be observed from the right portion of Fig.~\ref{fig:prototype}.

By conducting an offline gray-code pattern calibration, we can compute a homography between the camera and the projector.
We place visible patterned markers on our wearable headset, as shown in the left portion of Fig.~\ref{fig:prototype}.
As soon as markers on the wearable headset are visible to the camera through the steering mirror, we detect the markers to compute a homography between the screen and the camera.
We also extract the 2D offset from the projection center to the screen center for the tracking.
From the markers, we also calculate the distance from the projector to the screen by using the pose estimation function from OpenCV.
We get the focal length for the focus-tunable lens as the sum of the distance and the offset from the optical center of the projection optics to that of the camera.

For the steering in our proof-of-concept system, we implemented a naive sequential tracking algorithm. 
If the offset is exceeding a particular value, we steer the mirror towards that direction.
Although this sequential update does not precisely align the center of the projection and that of the target screen area, our homography mappings can absorb the remaining offset error.

To minimize the system latency as analyzed in Sec.~\ref{subsec:tracking_latency}, we implemented the above pipeline in three threads with C++.  We used ALIENWARE 15 (model 2019Q4, Core i9, Windows 10, and NVIDIA GTX1080).
The first thread captures the image from the camera.
The second thread detects AR markers using ArUco marker detection.
The third thread is responsible for 2D homography calculation, screen coordinate calculation (Perspective-n-Point method), 2D image warping, and image display on the projector.

\subsection{Passive wearable headset.}
We teardown a Lenovo Mirage Solo, an AR see-through headset, and harvest its bird-bath optics.
The bird-bath optics contains a beam splitting mirror and a beam-combiner per eye. 
The harvested beam-combiner can be observed as in the left portion of Fig.~\ref{fig:prototype}.
We install a rear-projection diffusive screen on top of the harvested bird-bath optics.
The active area of this diffusive screen is large as $30$~mm by $20$~mm per eye.
The active area on the diffusive screen here refers to regions that a user can see while wearing the headset.

We also place visible markers for tracking onto that diffusive screen.
Those visible markers are placed so that it does not overlap with the active area of that diffusive screen. We augment this entire assembly with 3D printed parts to turn it into a complete wearable headset. Finally, we add a mirror on top of the 3D printed parts, so that a projector sitting on a desk lower than the height of a headset can project images to the diffusive screen by bouncing the images off that mirror.
The added mirror can be completely omitted if the projector is installed at heights higher than the headset's height. 
The entire headset weighs $122$~g excluding a headband. The weight is smaller than the majority of smartphones ($150-200$~g).

Images that reaches to the diffusive screen are relayed to a user's eyes with the help of the beam splitters and the beam combiners found in the bird-bath optics.
We choose to use bird-bath optics as it's a common optical layout in near-eye displays.
To reduce the bulk of our headset system even further, our technique can potentially be used with other optical relays such as various kinds of optical waveguides and slimmer variants of bird-bath optics.

\section{Evaluation}\label{sec:evaluation}
Using our prototype, we analyze the proposed method with a series of experiments, and report our findings within this section.
In our next section, we will discuss means to improve our setup at a possible future.

\subsection{Image quality}\label{subsec:resolution_evaluation}
First, we investigate the resolution of our beaming display for different throw distances.
We choose to explore throw distances, $d_{image}$, of $0.5$~m, $1$~m, $1.5$~m, and $2.0$~m, which are likely distances for practical use cases.

\begin{figure}[t]
    \centering
    \includegraphics[width=0.9\linewidth]{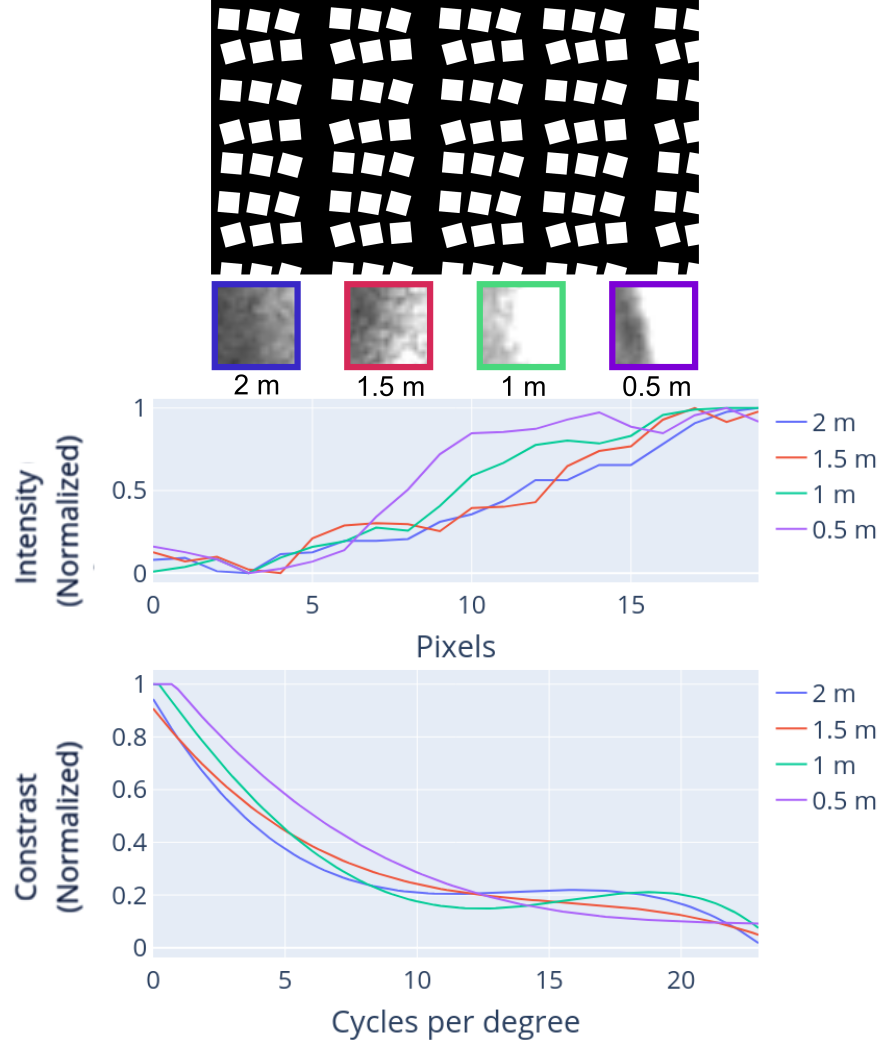}
    \vspace{-4mm}
    \caption{Resolution characteristics. Top: Test image used for slanted edge-based Modulation Transfer Function (MTF) analysis among sample slanted edge profiles from actual photographs captured. Middle: Edge profiles extracted from captured, sample slanted-edge photographs for $0.5$~m, $1.0$~m, $1.5$~m, and $2.0$~m throw distances. Bottom: Calculated MTFs as $3.0$~cycles per degree (cpd),$4.0$~cpd, $4.5$~cpd, and $7.0$~cpd at half contrast for throw distances of $0.5$~m, $1$~m, $1.5$~m, and $2.0$~m. Higher the cpd, the better the image resolution. The 20/20 vision in humans considered to be $30$~cpd.}
    \label{fig:mtf}
\end{figure}

For the quantitative assessment of the image resolution, we rely on a commonly accepted method, slanted edge Modulation Transfer Function (MTF) analysis~\cite{burns2000slanted}.
During our experimentation, we maintain a direct line of sight between the wearable headset and the steering projector.
The wearable headset is aligned with the angle of the steering projector, simulating a case where a user is looking towards the steering projector.
Using our steering projector, we project a test image shown in the top row of Fig.~\ref{fig:mtf}.
This specific test image contains boxes with slanted edges to help us capture edge profiles at different locations within the FoV of our wearable headset.

We capture photographs of the test images as seen through our wearable headset using Ximea MC023CG-SY-UB camera equipped with a $16$~mm Computar M1614-MP2 C-mount lens.
A human observer's eye can have an F-number in between 2.8 to 8, depending on the brightness levels of a scene.

During our capture, we set the F-number of our camera to $4$ to approximate a human observer's eye.
We crop regions of interest from captured photographs that contain slanted edges.
Some examples of such regions of interest are provided in the top row of Fig.~\ref{fig:mtf}.
We analyze the edge profiles from the regions of interest so that we get an idea about the variation from black pixels to white pixels, as shown in the middle plot in Fig.~\ref{fig:mtf}.
Finally, we compute the MTF of our prototype, and our findings suggest that, at $0.5$~m throw distance, the resolution of our system is 7~cycles per degree (cpd) with half contrast.

Current day, consumer-level near-eye displays typically provide resolutions in between $5-15$~cpd.
In this case, our proof-of-concept prototype promises resolutions matching a consumer-level VR near-eye display when short throw distances are the case.

In between $0.5$~m to $2.0$~m throw distances, there is a sharp drop in resolution from $7$~cpd to $3$~cpd.
The contrast of the projected images is inversely proportional to the throw distance; This sharp drop of the MTF is also related to the decrease in contrast as the projection distance increases.
A steering projector has to be equipped with a light source that can compensate for this brightness loss with the increasing throw distance.

\begin{figure*}[ht!]
    \centering
    \includegraphics[width=1.0\linewidth]{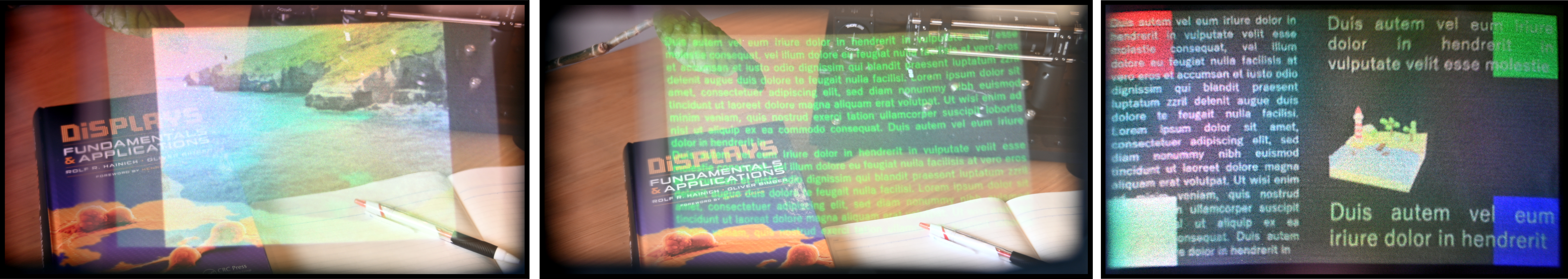}
    \vspace{-3mm}
    \caption{Photographs from a user's perspective. Left: Steering projector projects an image of a seaside view. Middle: Steering projector projects text, Right: steering projector projects images of a 3D model with some text and colored boxed at the edges, while the real-world view is blocked with an opaque material.}
    \label{fig:eyebox}
\end{figure*}

In our steering projector, projected pixel sizes are also increasing with the increasing throw distance.
Therefore, this sharp drop in resolution is also due to the need for an optical system that can project images with a fixed pixel size across all possible throw distances.
In our experiments, we observe that the resolution is homogeneous across different parts of the FoV, which can be observed from the sample photos in Fig.~\ref{fig:eyebox}.

With our wearable headset, when we set the throw distance to $2$~m, the monocular FoV can be as large as $36$~degrees along the axis that goes from left to right of a user and $24$~degrees along the axis that goes from top to bottom.
In our prototype, our monocular FoV with $0.5$~m throw distances shrinks to $24$ by $17$~degrees.
We use an off-the-shelf near-eye display optics.
However, different relay optics readily available on the market can be used as the eyepiece of our wearable headset to improve both the resolution and FoV characteristics of our prototype.

\subsection{Tracking and latency} \label{subsec:evaluation_tracking}

\subsubsection{Tracking space}
In a beaming display, a volume that represents possible user locations with possible head orientations can be described as a projection volume.
We design to allow head rotations of $20$~degrees along the axis that goes from left to right of a user,  and  $20$~degrees along the axis that goes from left to right of a user.
In our prototype, we allow throw distances as large as $2.0$~m, and our steering projector can project images in a cone with $30$ by $30$ degrees.
All of the constraints in head rotations, throw distances and scan cones are in place to keep the prototype work with a reasonable performance.

\subsubsection{Tracking functions} 
\begin{figure}[t]
    \centering
    \includegraphics[width=\linewidth]{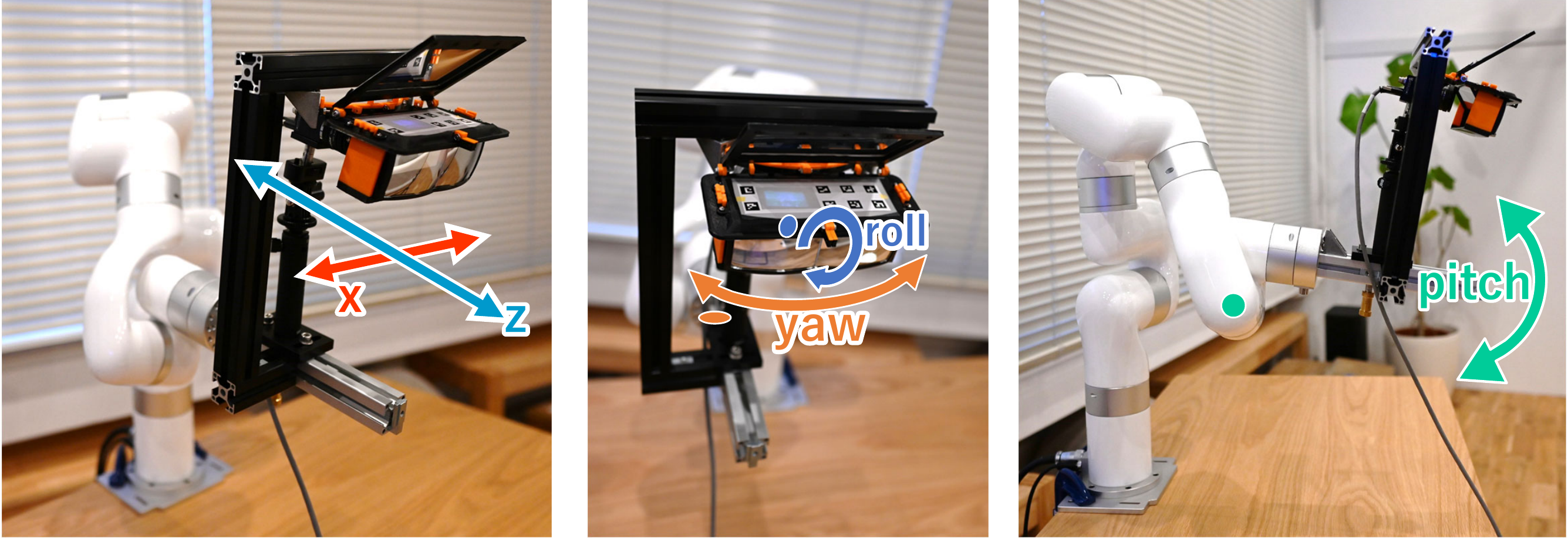}
    \vspace{-5mm}
    \caption{Translation and rotation references of the robot arm with a wearable headset. These references are used in the evaluation in Sec.~\ref{subsec:evaluation_tracking}.}
    \label{fig:tracking-setup}
\end{figure}

\begin{figure*}[ht!]
    \centering
    \includegraphics[width=0.85\linewidth]{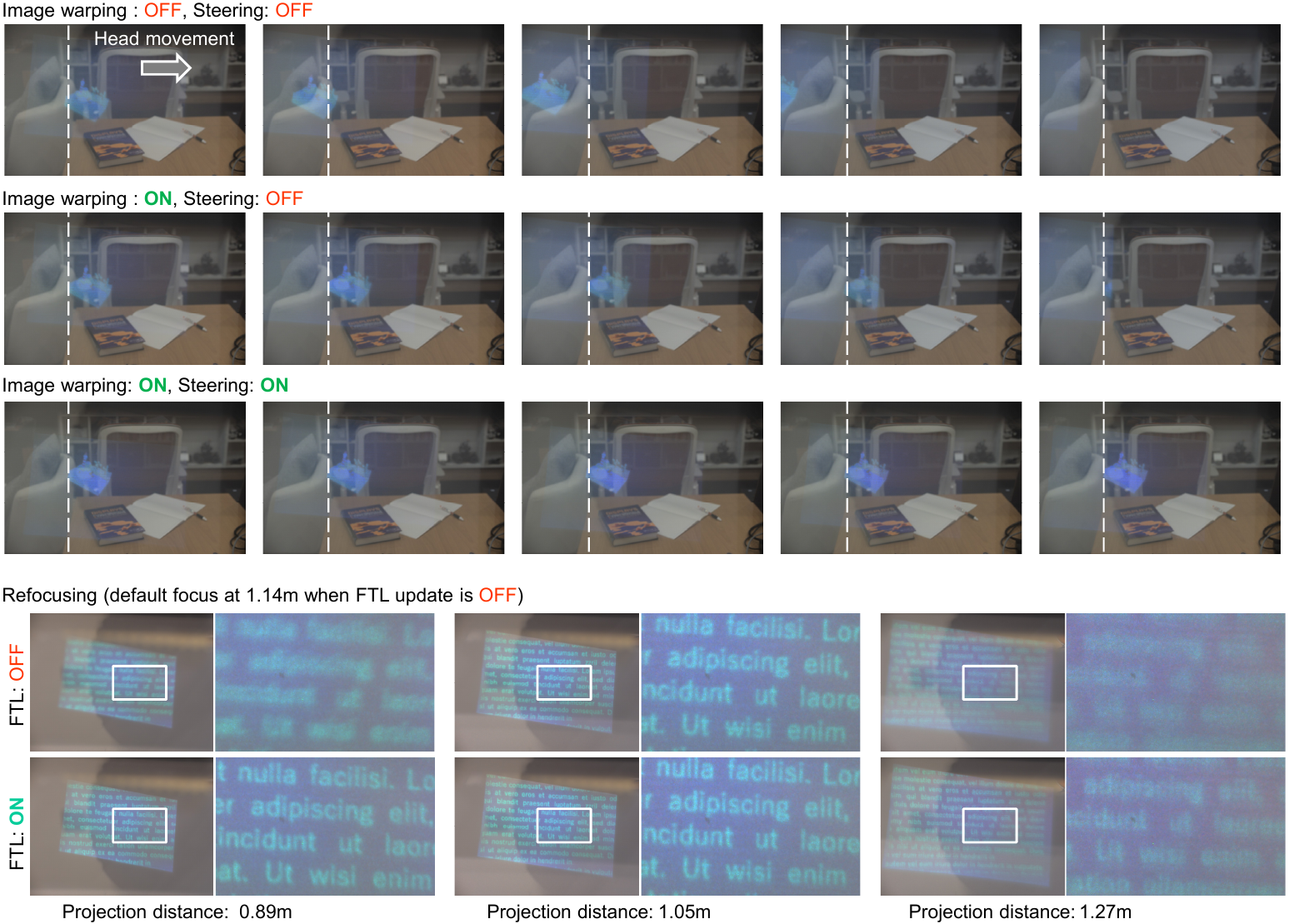}
    \vspace{-3mm}
    \caption{Evaluation of the tracking functions. Top: demonstrating the effect of the image warping and the beam steering. A robot arm moved the headset. Bottom: Demonstrating the effect of the projector refocusing with the focus-tunable lens (FTL). The top row is views when the FTL lens focuses at 1.14m. The bottom row shows views when the FTL dynamically refocuses on the changing projection distance.}
    \label{fig:tracking-result}
\end{figure*}

Figure~\ref{fig:tracking-result} demonstrates how our tracking unit helps to realize the desired projection on the screen of the wearable screen. Pleasec onsult with our supplemental videos for more detail.
To quantitatively simulate head motion, we mount both the headset and the user-perspective camera on a robot arm and capture the user view for each head pose. 
The arm we used is a UFACTORY xArm 7, a 7-axis arm with the position repeatability of 0.1mm (Fig.~\ref{fig:tracking-setup}). 

We first demonstrate the effect of image warping and beam steering.
We use the x-axis motion shown in Fig.~\ref{fig:tracking-setup}.
When both the image warping and the steering are off, the system is simply a static projector. The view is only valid when the headset is at the initial position (Fig.~\ref{fig:tracking-result} Top first raw).
When the image perspective is on while steering is off, the system can project the correct perspective at the beginning. While the head position shifts by a small amount, however, the projection is quickly out of the field of view, and the projected image is cutoff (Fig.~\ref{fig:tracking-result} Top second raw).
Finally, When both the steering and the image warping are activated, the rendered images stay fixed even the viewpoint moves. 

Since the projector has a single focus depth at the time, we also need to tune the focus-tunable lens~\cite{iwai2015extended}. Fig.~\ref{fig:tracking-result} bottom shows qualitative results of the effect of the focus-tunable lens. 
We captured the projection at three different projection distances (0.89 m, 1.05 m, and 1.27 m).
For each capture, we used a default focus parameter 1.14 m first, then turn on the depth tracking to dynamically tune the focus-tunable lens.
As seen in the second row in the figure, dynamic refocusing is necessary to maintain the image quality.

\subsubsection{System latency}\label{subsec:tracking_latency}
Quantitatively analyzing the latency of an AR system is practically essential. AR systems with tracking are inherently causal since displayed image is generated based on the latest tracking measurement. Increasing the frame rate will often lead to a reduction in latency.


To achieve a reasonable performance in terms of the latency, we design the pipeline of our proof-of-concept system to suppress latency as much as we can.
Several components may have different types of latency. They include the mirror, the camera, the projector, and their communication protocols.

The Optotune's steering mirror accepts commands via SPI connection at up to 10 kHz. The connected development board (Nuclero 144) has SPI ports with a clock speed of 50 MHz. The SPI clock of the mirror driver is 4 Mbps. Thus we can safely assume that the communication with the mirror is running at an ideal speed. 
For the tracking pipeline with the steering projector, as mentioned in Sec.~\ref{subsec:implementation:tracking_unit}, the pipeline is divided into three parts: capture, tracking, and steering and display. Note that, for the tracking evaluation, we swapped the tracking and the user-view cameras so that we can use a camera with a higher maximal frame rate (Ximea MC023CG-SY-UB, max. 165fps). The 1st thread for image capture ran at about 130fps. The thread resized the image to half. The 2nd thread for the marker detection ran at 130 fps. The 3rd thread for the steering and image display ran at 100 fps.

\begin{figure}[t]
    \centering
    \includegraphics[width=\linewidth]{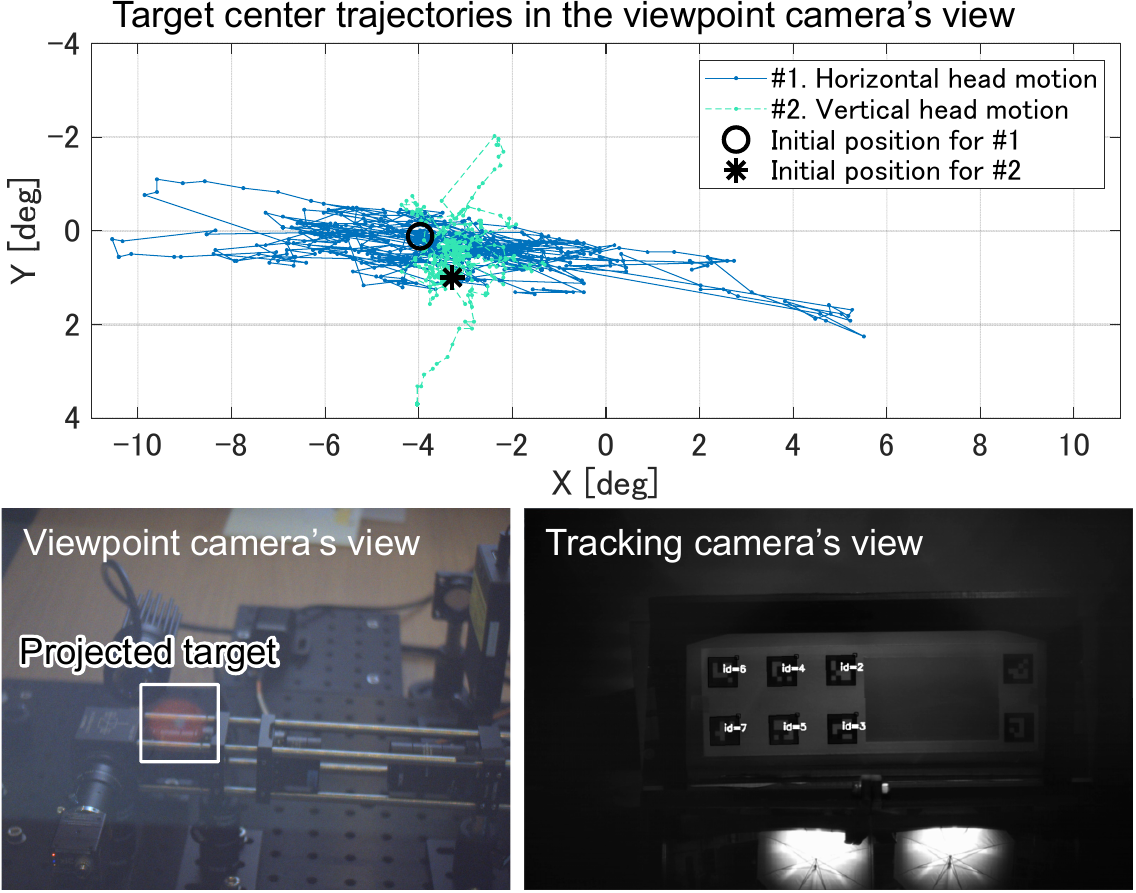}
    \vspace{-5mm}
    \caption{Latency evaluations. Top: Trajectories of the projected image's center in the viewpoint coordinates under two head motions. Bottom left: The viewpoint camera’s view at the initial position of the vertical head motion. A sample capture of the steering tracking camera where several markers are detected.
    The viewpoint camera had a resolution of 1032x772 pixels.}
    \label{fig:projected_center}
\end{figure}

For the latency evaluation, we assumed a seated situation where the user is working in an office environment. In such a scenario, the head and torso movements are dominant motion factors~\cite{sidenmark2019eye}. Following the work by Sidenmark and Gellersen~\cite{sidenmark2019eye}, we defined an ordinary rotation speed by head and torso to be 20 deg/sec. We rotated the headset mounted on the robot arm in both horizontal and vertical directions with the defined speed and recorded the viewpoint view. During these recording, the viewpoint camera observes a target projected on the screen. We chose a white dot surrounded with a red circle as a target to make it easy to be detected for analysis. We also used a studio light to stabilize the marker tracking. Fig.~\ref{fig:projected_center} shows trajectories of the center of the projected image manually labeled from recorded videos. The vertical motion resulted in smaller tracking jitters than the horizontal motion. 

Our observation shows that these differences are caused by the head pose tracking instability from the AR marker detection. This implies that head tracking mechanism has to be improved to overcome the observed jitters. Subjective observation by a real user taking the headset also revealed that the tracking range is limited due to the current marker layout, thus improving the tracking design is necessary.

While the source of the remaining latency stems from several factors, the current proof-of-concept system's major factor is the marker-based tracking due to its high computational cost and low stability of the pose estimation. As an alternative tracking option for high-speed applications, we can ideally replace the tracking module with an infrared marker-based system with a high-speed camera~\cite{mikawa2018variolight}. We discuss further in this potential research area in Sec.~\ref{sec:discussion}. Please consult to our supplementary  material for the video recordings of our latency experiment.



\subsubsection{Image appearance}
We evaluate the image appearance of the beaming display when the user moves in the working area. 
We employ basic translation and orientation motion using the robot arm jig as described before (Fig.~\ref{fig:tracking-setup}).
Note that we capture each image as a still capture taken once the tracking is stabilized for given poses.
By doing this way, we can assess the quality limit of the beaming display under motions typical in an office environment.

The motion sequences we tested in the evaluation are two linear motions (slide and depth) and three orientation motions (yaw, pitch, and roll).
Please consult to our supplementary material for the video recordings of our experiment.
The yaw motion would be the most typical head orientation when the user is working on a desk environment. Pitch corresponds to motion when the user is looking down from the scene to the desk. 

\begin{figure*}[ht!]
    \centering
    \includegraphics[width=0.85\linewidth]{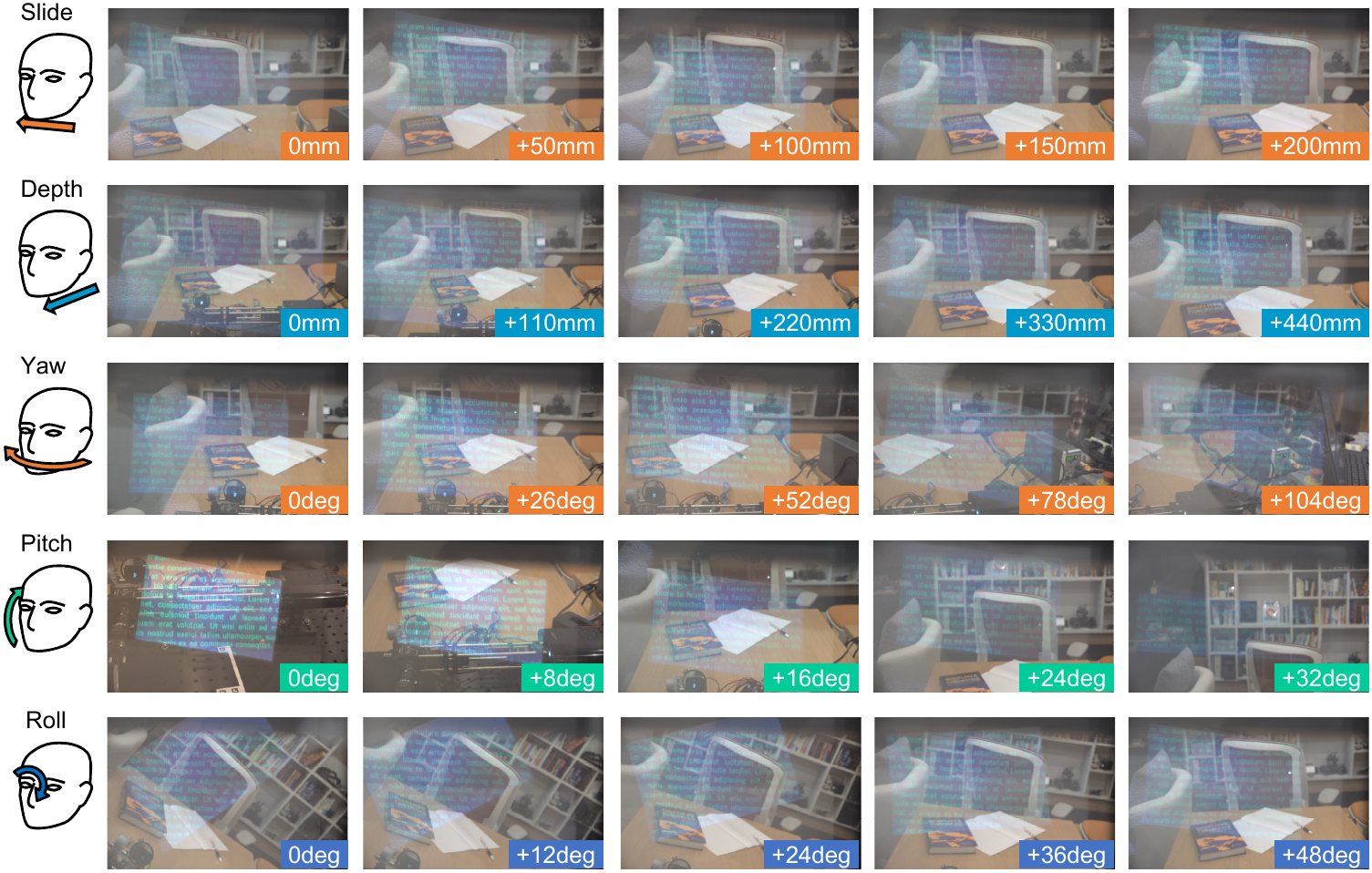}
    \vspace{-3mm}
    \caption{Qualitive demonstration of the image quality in tracking scenarios with basic head motion sequences. Lengths and angles represent the amount of corresponding motion as offsets. Note that the image projected on the screen keeps the same position except for the visible area due to steered projection.}
    \label{fig:tracking-test-result-images}
\end{figure*}

Figure~\ref{fig:tracking-test-result-images} summarizes our findings for image appearance experiment. 
For the horizontal slide motion, the projected image does not change much.
For the motion along with the depth, the FoV of the projection in the view changes, leading to the change of the resolution as elaborated in Sec.~\ref{subsec:resolution_evaluation}.
For the yaw motion, the characteristic of the projection is similar to the horizontal translation, yet the change of the yaw rotates projected image FoV.
For the pitch motion, the characteristic of the projection is similar to the horizontal translation. Yet, the angle of the receiver screen against the projection axis gets eccentric, and the projected image is easier to be cropped.
For the roll motion, the roll angle directly affects the angle of the projection FoV.
Note that the projection area on the diffusive surface is also rotating in shape as we translate and rotate a user's view.
We believe image rotators~\cite{miyashita2015high} can be integrated into our optical system to dynamically rotate the shape of the projected on a diffusive screen.

\section{Discussion and Future work}\label{sec:discussion}
We demonstrate a prototype for our beaming display proposal. 
We believe this prototype and the method can trigger interesting research questions in the near future in the fields of optical design, tracking systems, computational hardware, and human vision.
Hence, we next provide a discussion of the important aspects of our design and identify a set of future research directions.

\paragraph{Resolution.}
With our prototype, we demonstrate resolutions that are matching a consumer-level grade VR headsets. 
In our steering projector prototype, we use a pico projector with 854 by 480 pixels. There are currently projectors in the market with much higher resolutions (\ie, 8K resolutions).
Switching to a higher resolution projector can potentially improve the perceived resolution of a beaming display. At the same time, there are also techniques to improve the resolution of projectors beyond what is available on the market~\cite{akcsit2020patch}.
A dedicated projector design with freeform optical design approaches~\cite{bauer2018starting} can potentially help to improve the resolution performance of the optical system used in a steering projector.
Hence, exploring the design of more effective optical components for beaming displays is an interesting future direction. 
Besides, we are also planning to merge holographic approaches that take advantage of speckle fields~\cite{kuo2020high} with beaming displays, and explore possible resolution benefits of holographic techniques in large throw distance projections.

\paragraph{Tracking and latency.}
For stable AR rendering on beaming displays, the steered projection has to be as accurate as possible with low latency.
The image projected on the screen appears magnified to the user due to the bird-bath optics. 
While this increases the image FoV, it also amplifies the projection error, because a slight change of the projection position appears as a larger shift in the user's field of view. 
Fortunately, the tracking camera is co-axial with the projector, so the image warping can accommodate the steering error by compensating the image shift.

A sophisticated image warping based on head orientation speed can improve the perceived latency, yet requires fast head motion detection that is typically achieved by rotation encoding mechanism installed on the headset~\cite{lincoln2016motion}.
A high-speed projection mapping system is another engineering option to tackle this latency issue~\cite{wang2020high}.
Infrared (IR) markers are typically used to avoid visually occupying visible markers~\cite{narita2016dynamic}.

As we discussed in Sec.~\ref{subsec:tracking_unit}, there are several options for steering mirrors.
A typical choice in steerable projection is a galvanometer optical scanner~\cite{narita2016dynamic}. 
Commercial products have small angler repeatability, such as 15$\mu$rad (\eg, Thorlabs GVS001).
The voice coil mirror from Optotune we used has a range of 30-100$\mu$rad error.
Base on the simulation setup above, these specifications lead to 0.003mm or 0.006-0.02mm image errors at the screen 1m away with a $45$deg projection angle.
The errors are equal to 0.015\% or 0.03-0.1\% of the screen image size.

The overall error of the projection is, however, the sum of all other errors. The errors include screen detection by the scene camera and the latency of the rendering. Other error sources are the tracking accuracy of the screen in the scene camera and the overall system latency.

For the scene camera, marker-based tracking has various options.
Examples include IR tracking with retroreflective markers that can typically be faster than visual marker tracking.
The scene camera itself has the focal length; thus, tracking might be unstable if the tracking area is long in the optical axis direction.

The latency of the system may significantly impact the visual performance.
We demonstrated the potential of the beaming display with the proof-of-concept system with compelling image quality and latency. Our system still has a room to be optimized for the tracking speed.
Existing works prove that the optimization of the projection speed is possible by combining a more sophisticated high-speed tracking and projection framework~\cite{narita2016dynamic,lincoln2016motion}.

The head motion is another factor that induces a projection error on top of the latency issue.
The requirement for tracking is affected by involuntary human head movements. For example, when working in a desk environment, the involuntary head movements of a user focusing on a computer screen tend to be suppressed more when the cognitive load increases~\cite{dirican2012involuntary}. Because the user's eyes, head, and torso coordinate each other's movements, head movements may also be relatively limited since gaze motion can account for a cognitive task such as visual targeting~\cite{sidenmark2019eye}.

\paragraph{Projection volume and number of users.}
The projection volume is bounded as a line of sight is required in between a steering projector and a wearable headset.
Therefore, interactions requiring users to rotate their heads away from the projector would not be possible with our current beaming display prototype.
The current beaming display prototype demonstrated in this paper also supports only a single user at a time.
According to survey from a VR headset manufacturer\footnote{https://steamcommunity.com/app/358720/discussions/0/ \\ 350532536103514259/?ctp=2\#c133258092253222557}, most users use $1$m by $1$m area during usage, and a small portion of the users use $3$m by $2.5$m area, which is on par with our measured projection volume of $2$m by $2$m that provides a reasonable image quality.

Following the spirit of the work by Kawahara et al.~\cite{kawahara2018dynamic}, to increase a projection volume, steering projectors can be mounted on a moving body such as a drone or robot. They can be relocated according to a user's needs.

To enable large projection volumes and to support a larger number of users in the future, we need to improve the current system. Such improvements include: multiple steering projectors per eye or user, active mirrors distributed in the environment to steer the projected beams towards a user from different angles coupled with faster steering mirrors

\paragraph{Miniaturization of passive wearable headset.}
The wearable headset in our prototype is based on a bird-bath optics approach. 
We have built our headset by harvesting optical components from an existing consumer-level headset from a previous generation.
Modern variants of bird-bath optics on the near-eye display market are closely approximating the form of a pair of sunglasses (i.e., Nreal).
Switching to smaller optics also requires redesigning steering projector optics because the diffusive screen area in a smaller headset will also be smaller in size.

As a more advanced approach, the work by Kuo et al.~\cite{kuo2020high} proposes a holographic approach relying on patterned diffusers that can be used as an eyepiece without requiring any additional lenses or mirrors. 
In the near future, we plan on expanding the work by Kuo et al.~\cite{kuo2020high} by switching to a holographic projection mechanism and a patterned diffuser used as an eyepiece instead of bird-bath optics.

\paragraph{Ambient and projector light leakage.}
Even though our prototype projects an image all filled with black pixels, viewers can observe an unintended background, which can also be observed as a user's point of view photograph in Fig.~\ref{fig:teaser}.
This constant background is causing a loss of contrast in the observed image, and it is a combined result of two different factors that we have identified.
The first factor is the ambient light reaching on the diffusive screen of our wearable headset, and the second factor is the black levels of our projector.

The ambient light leakage can be decreased by using a notch filter on the wearable headset to let only the certain colors of light reach on the diffusive screen.
The notch filter approach would also require to redesign the projector's light engine with specific color primaries.
In fact, consumer-level display systems that use notch filters and dedicated projectors were built in the past for 3D displays with glasses~\cite{woods20093}.
Even further, to completely avoid the light from ambient reaching at the diffusive screen, the light sources that are present in the ambient can be replaced with the ones that do not contain color primaries used in such a projector.

To overcome the limitation of the projector's black levels, the contrast level of the projector has to improve. In our case, our prototype uses a LED illuminated SLM with a poor contrast ratio of $2000:1$.
In the next iterations, we plan on switching to other projectors with much higher contrast ratios as they are readily available on the market.

\paragraph{Privacy.}
Beyond technical issues, obstacles in a social context such as acceptance or privacy concerns regarding cameras on AR near-eye displays pose a major challenge for the adoption of various kinds of AR display technologies.
Though we didn't conduct a formal subjective test on acceptance of our prototype, we expect that placing cameras and sensors away from a user, as in our case, may have a positive impact from the perspective of a user.

\paragraph{Classical near-eye displays vs beaming display}
Similar to the current eco-system of the displays, multiple types of displays such as 3D displays and near-eye displays are expected to co-exist in the future. 
Beaming display as a partial variant of AR near-eye displays will co-exist alongside AR near-eye displays, and we expect AR near-eye display to be miniaturized further.
Unlike AR near-eye displays, though, each part of a beaming display, such as the projector, glasses, or tracking unit, can be upgraded in a stand-alone way without having to change the hardware as a whole completely.
The design approach of beaming displays leaves room for a modular design while enabling large computational resources as a steering projector can be connected to such resources without having to suffer from any miniaturization, heat, or power-related issues.

\paragraph{Applications.}
Our approach can enable long-duration usage of AR applications. 
Given the current conditions with the pandemic at the time of this paper, virtualization is on an increasing demand with cases such as remote-work and teleconference.
Enabling improved long duration usage in hardware can potentially help to improve adoption of AR tools in the long run, and can support the trend of virtualization, greatly.
Therefore, we believe that, in this new era, our method is potentially useful for improving teleconferencing, remote-work, education, gaming, and creative design.

\section{Conclusion}\label{sec:conclusion}
AR offers an attractive future, where computer-generated visuals improve our daily lives and routines when and where it is needed.
Towards that future, AR near-eye displays have to be free from any heating problems, form-factor, and weight-related issues or computational and power issues due to limited onboard resources.
Our work proposes a new class of AR near-eye displays that can potentially help with issues related to ergonomics, computation, and power.
As a proof-of-concept, we demonstrate a steering projector with tracking cameras and an all passive wearable headset prototype.
We evaluate and discuss our prototype and demonstrate resolutions matching current consumer-level headsets.
Though some challenges remain, we believe that novel variants of our method can enable an all-day wearable AR display of the future.

\acknowledgments{
The authors are thankful and grateful to Duygu Ceylan, Daisuke Iwai, Toshiyuki Amano, and Kiyoshi Kiyokawa for the fruitful discussions.
Yuta Itoh are partially supported by JST PRESTO Grant Number JPMJPR17J2 and JSPS KAKENHI Grant Number and JP20H04222, Japan. 
}


\bibliographystyle{abbrv-doi}

\bibliography{ref}

\begin{thebibliography}{10}

\bibitem{akcsit2020patch}
K.~Ak{\c{s}}it.
\newblock Patch scanning displays: spatiotemporal enhancement for displays.
\newblock {\em Optics Express}, 28(2):2107--2121, 2020.

\bibitem{akcsit2019manufacturing}
K.~Ak{\c{s}}it, P.~Chakravarthula, K.~Rathinavel, Y.~Jeong, R.~Albert,
  H.~Fuchs, and D.~Luebke.
\newblock Manufacturing application-driven foveated near-eye displays.
\newblock {\em IEEE transactions on visualization and computer graphics},
  25(5):1928--1939, 2019.

\bibitem{bauer2018starting}
A.~Bauer, E.~M. Schiesser, and J.~P. Rolland.
\newblock Starting geometry creation and design method for freeform optics.
\newblock {\em Nature communications}, 9(1):1--11, 2018.

\bibitem{benko2015fovear}
H.~Benko, E.~Ofek, F.~Zheng, and A.~D. Wilson.
\newblock Fovear: Combining an optically see-through near-eye display with
  projector-based spatial augmented reality.
\newblock In {\em Proc. of the 28th ACM UIST}, pp. 129--135, 2015.

\bibitem{benko2014dyadic}
H.~Benko, A.~D. Wilson, and F.~Zannier.
\newblock Dyadic projected spatial augmented reality.
\newblock In {\em Proceedings of the 27th annual ACM symposium on User
  interface software and technology}, pp. 645--655, 2014.

\bibitem{bimber2006modern}
O.~Bimber and R.~Raskar.
\newblock Modern approaches to augmented reality.
\newblock In {\em ACM SIGGRAPH 2006 Courses}, pp. 1--es. 2006.

\bibitem{burns2000slanted}
P.~D. Burns et~al.
\newblock Slanted-edge mtf for digital camera and scanner analysis.
\newblock In {\em Is and Ts Pics Conference}, pp. 135--138. SOCIETY FOR IMAGING
  SCIENCE \& TECHNOLOGY, 2000.

\bibitem{de2013progress}
J.~De~Smet, A.~Avci, P.~Joshi, D.~Schaubroeck, D.~Cuypers, and H.~De~Smet.
\newblock Progress toward a liquid crystal contact lens display.
\newblock {\em Journal of the Society for Information Display}, 21(9):399--406,
  2013.

\bibitem{dirican2012involuntary}
A.~C. Dirican and M.~G{\"o}kt{\"u}rk.
\newblock Involuntary postural responses of users as input to attentive
  computing systems: An investigation on head movements.
\newblock {\em Computers in Human Behavior}, 28(5):1634--1647, 2012.

\bibitem{foreman2011computational}
M.~R. Foreman and P.~T{\"o}r{\"o}k.
\newblock Computational methods in vectorial imaging.
\newblock {\em Journal of Modern Optics}, 58(5-6):339--364, 2011.

\bibitem{hamasaki2019varifocal}
T.~Hamasaki and Y.~Itoh.
\newblock Varifocal occlusion for optical see-through head-mounted displays
  using a slide occlusion mask.
\newblock {\em IEEE TVCG}, 25(5):1961--1969, 2019.

\bibitem{hedili2013microlens}
M.~K. Hedili, M.~O. Freeman, and H.~Urey.
\newblock Microlens array-based high-gain screen design for direct projection
  head-up displays.
\newblock {\em Applied optics}, 52(6):1351--1357, 2013.

\bibitem{herpich2017comparative}
F.~Herpich, R.~L.~M. Guarese, L.~M.~R. Tarouco, et~al.
\newblock A comparative analysis of augmented reality frameworks aimed at the
  development of educational applications.
\newblock {\em Creative Education}, 8(09):1433, 2017.

\bibitem{hsu2014transparent}
C.~W. Hsu, B.~Zhen, W.~Qiu, O.~Shapira, B.~G. DeLacy, J.~D. Joannopoulos, and
  M.~Solja{\v{c}}i{\'c}.
\newblock Transparent displays enabled by resonant nanoparticle scattering.
\newblock {\em Nature communications}, 5(1):1--6, 2014.

\bibitem{iwai2015extended}
D.~Iwai, S.~Mihara, and K.~Sato.
\newblock Extended depth-of-field projector by fast focal sweep projection.
\newblock {\em IEEE transactions on visualization and computer graphics},
  21(4):462--470, 2015.

\bibitem{kade2015head}
D.~Kade, K.~Ak{\c{s}}it, H.~{\"U}rey, and O.~{\"O}zcan.
\newblock Head-mounted mixed reality projection display for games production
  and entertainment.
\newblock {\em Personal and Ubiquitous Computing}, 19(3-4):509--521, 2015.

\bibitem{kaminokado2020stainedview}
T.~Kaminokado, Y.~Hiroi, and Y.~Itoh.
\newblock Stainedview: Variable-intensity light-attenuation display with
  cascaded spatial color filtering for improved color fidelity.
\newblock In {\em IEEE TVCG}, pp. (3576--3586, 2020.

\bibitem{kawahara2018dynamic}
T.~Kawahara, D.~Iwai, and K.~Sato.
\newblock Dynamic path planning of flying projector considering collision
  avoidance with observer and bright projection.
\newblock In {\em Proceedings of the 23rd International Conference on
  Intelligent User Interfaces Companion}, pp. 1--2, 2018.

\bibitem{Koulieris2019}
G.~A. Koulieris, K.~Ak{\c{s}}it, M.~Stengel, R.~K. Mantiuk, K.~Mania, and
  C.~Richardt.
\newblock Near-eye display and tracking technologies for virtual and augmented
  reality.
\newblock In {\em Computer Graphics Forum}, vol.~38, pp. 493--519. Wiley Online
  Library, 2019.

\bibitem{kress2019optical}
B.~C. Kress.
\newblock Optical waveguide combiners for ar headsets: features and
  limitations.
\newblock In {\em Digital Optical Technologies 2019}, vol. 11062, p. 110620J.
  International Society for Optics and Photonics, 2019.

\bibitem{kuo2020high}
G.~Kuo, L.~Waller, R.~Ng, and A.~Maimone.
\newblock High resolution {\'e}tendue expansion for holographic displays.
\newblock {\em ACM Transactions on Graphics (TOG)}, 39(4):66--1, 2020.

\bibitem{lan2019metasurfaces}
S.~Lan, X.~Zhang, M.~Taghinejad, S.~Rodrigues, K.-T. Lee, Z.~Liu, and W.~Cai.
\newblock Metasurfaces for near-eye augmented reality.
\newblock {\em ACS Photonics}, 6(4):864--870, 2019.

\bibitem{langlotz2018chroma}
T.~Langlotz, J.~Sutton, S.~Zollmann, and H.~Itoh, Yuta\&~Regenbrecht.
\newblock Chromaglasses: Computational glasses for compensating colour
  blindness.
\newblock In {\em Proceedings of the 2018 CHI Conference on Human Factors in
  Computing Systems}, pp. 390:1--390:12. ACM, New York, NY, USA, 2018.

\bibitem{likamwa2014draining}
R.~LiKamWa, Z.~Wang, A.~Carroll, F.~X. Lin, and L.~Zhong.
\newblock Draining our glass: An energy and heat characterization of google
  glass.
\newblock In {\em Proceedings of 5th Asia-Pacific Workshop on Systems}, pp.
  1--7, 2014.

\bibitem{lincoln2016motion}
P.~Lincoln, A.~Blate, M.~Singh, T.~Whitted, A.~State, and H.~Lastra,
  Anselmo\&~Fuchs.
\newblock From motion to photons in 80 microseconds: Towards minimal latency
  for virtual and augmented reality.
\newblock {\em IEEE Trans. on Visualization and Computer Graphics},
  22(4):1367--1376, 2016.

\bibitem{martin2020mojo}
P.~S. Martin.
\newblock Mojo vision nanoleds for invisible computing.
\newblock In {\em Light-Emitting Devices, Materials, and Applications XXIV},
  vol. 11302, p. 1130204. International Society for Optics and Photonics, 2020.

\bibitem{mikawa2018variolight}
Y.~Mikawa, T.~Sueishi, Y.~Watanabe, and M.~Ishikawa.
\newblock Variolight: Hybrid dynamic projection mapping using high-speed
  projector and optical axis controller.
\newblock In {\em SIGGRAPH Asia 2018 Emerging Technologies}, pp. 1--2. 2018.

\bibitem{miyashita2015high}
L.~Miyashita, Y.~Watanabe, and M.~Ishikawa.
\newblock High-speed image rotator for blur-canceling roll camera.
\newblock In {\em 2015 IEEE/RSJ International Conference on Intelligent Robots
  and Systems (IROS)}, pp. 6047--6052. IEEE, 2015.

\bibitem{mohan2009bokode}
A.~Mohan, G.~Woo, S.~Hiura, Q.~Smithwick, and R.~Raskar.
\newblock Bokode: imperceptible visual tags for camera based interaction from a
  distance.
\newblock In {\em ACM SIGGRAPH 2009 papers}, pp. 1--8. 2009.

\bibitem{narita2016dynamic}
G.~Narita, Y.~Watanabe, and M.~Ishikawa.
\newblock Dynamic projection mapping onto deforming non-rigid surface using
  deformable dot cluster marker.
\newblock {\em IEEE transactions on visualization and computer graphics},
  23(3):1235--1248, 2016.

\bibitem{okumura2012lumipen}
K.~Okumura, H.~Oku, and M.~Ishikawa.
\newblock Lumipen: Projection-based mixed reality for dynamic objects.
\newblock In {\em 2012 IEEE International Conference on Multimedia and Expo},
  pp. 699--704. IEEE, 2012.

\bibitem{Rekimoto1995}
J.~Rekimoto and K.~Nagao.
\newblock The world through the computer: Computer augmented interaction with
  real world environments.
\newblock In {\em Proceedings of the 8th annual ACM symposium on User interface
  and software technology}, pp. 29--36. ACM, 1995.

\bibitem{rolland2012see}
J.~P. Rolland, K.~P. Thompson, H.~Urey, and M.~Thomas.
\newblock See-through head worn display (hwd) architectures., 2012.

\bibitem{sidenmark2019eye}
L.~Sidenmark and H.~Gellersen.
\newblock Eye, head and torso coordination during gaze shifts in virtual
  reality.
\newblock {\em ACM Transactions on Computer-Human Interaction (TOCHI)},
  27(1):1--40, 2019.

\bibitem{soomro2016design}
S.~R. Soomro and H.~Urey.
\newblock Design, fabrication and characterization of transparent
  retro-reflective screen.
\newblock {\em Optics express}, 24(21):24232--24241, 2016.

\bibitem{sutherland1968head}
I.~E. Sutherland.
\newblock A head-mounted three dimensional display.
\newblock In {\em Proceedings of the December 9-11, 1968, fall joint computer
  conference, part I}, pp. 757--764, 1968.

\bibitem{wang2020high}
L.~Wang, H.~Xu, S.~Tabata, Y.~Hu, Y.~Watanabe, and M.~Ishikawa.
\newblock High-speed focal tracking projection based on liquid lens.
\newblock In {\em Special Interest Group on Computer Graphics and Interactive
  Techniques Conference Emerging Technologies}, pp. 1--2, 2020.

\bibitem{woods20093}
A.~Woods.
\newblock 3-d displays in the home.
\newblock {\em Information Display}, 25(7):8--12, 2009.

\bibitem{yamamoto201616}
A.~Yamamoto, Y.~Yanai, M.~Nagai, R.~Suzuki, and Y.~Ito.
\newblock 16-3: A novel transparent screen using cholesteric liquid crystal
  dots.
\newblock In {\em SID Symposium Digest of Technical Papers}, vol.~47, pp.
  185--188. Wiley Online Library, 2016.

\end{thebibliography}
\end{document}